\documentclass[floats,floatfix,showpacs,amssymb,prd,superscriptaddress,nofootinbib,nolongbibliography,reprint]{revtex4-1}

\usepackage{amssymb,amsmath,verbatim,mathtools,needspace,enumitem,etoolbox,graphicx,physics,microtype,afterpage,bigints,gensymb}

\usepackage{makecell}
\usepackage[dvipsnames, usenames]{xcolor}
\definecolor{linkcolor}{rgb}{0.0,0.3,0.5}
\definecolor{dodgerblue}{HTML}{1E90FF}
\usepackage[unicode, colorlinks=true, linkcolor=linkcolor, citecolor=linkcolor, filecolor=linkcolor,urlcolor=linkcolor, pdfusetitle]{hyperref}
\usepackage[all]{hypcap}
\usepackage[T1]{fontenc}
\usepackage[utf8]{inputenc}
\usepackage{orcidlink}
\usepackage[htt]{hyphenat}
\usepackage{graphicx}
\usepackage{booktabs} \usepackage{multirow}
\usepackage{rotating}
\usepackage{tikz}

\def\pymcpop{\texttt{pymcpop-gw} }

\makeatletter
\newcommand*{\balancecolsandclearpage}{\close@column@grid \cleardoublepage \twocolumngrid}	
\makeatother

\usepackage{lipsum}

\newcommand{\mrs}{\affiliation{Aix-Marseille Universit\'e, Universit\'e de Toulon, CNRS, CPT, Marseille, France}}

\begin{document}

\title{
Gravitational-wave constraints on $H_0$ are robust to putative redshift evolution in the binary black hole mass spectrum at current sensitivity
}

\author{Alessandro Agapito\texorpdfstring{$\,$}{}\orcidlink{0009-0005-9004-3163}}
\email{alessandro.AGAPITO@univ-amu.fr}
\mrs

\author{Viola De Renzis\texorpdfstring{$\,$}{ }\orcidlink{0000-0001-7038-735X}}
\email{viola.DE-RENZIS@univ-amu.fr}
\mrs

 \author{Michele Mancarella\texorpdfstring{$\,$}{ }\orcidlink{0000-0002-0675-508X}}
 \mrs


\date{\today}

\begin{abstract}
Spectral-siren cosmology constrains the Hubble constant $H_0$ using gravitational-wave observations of compact-binary coalescences. The method combines luminosity distances inferred from the waveform with redshift information statistically encoded in population features of the source-frame mass spectrum. Because the detector measures redshifted masses, structure in the intrinsic mass distribution acts as an internal ``ruler,'' making the inference sensitive to assumptions about the population model. In particular, redshift evolution of the mass spectrum is widely discussed as a potential systematic for $H_0$ measurements.
We revisit spectral-siren constraints with the GWTC-4.0 binary black hole catalog, explicitly allowing the main mass scales of a standard parametric mass model to evolve with redshift. We find no compelling evidence for evolution at current sensitivity. Allowing evolution produces a modest, nonstatistically significant shift of the $H_0$ posterior toward lower values, which we interpret with targeted posterior and event-level diagnostics. Importantly, the associated systematic uncertainty is subdominant to that induced by alternative redshift-independent descriptions of the mass spectrum, such as the number of spectral features and the functional form used to model them. Our results indicate that, at current sensitivity, spectral-siren constraints on $H_0$ are robust to redshift evolution of the mass spectrum within the flexibility explored here.
Using injection studies, we show that this mild $H_0$ shift is reproduced when a nonevolving underlying population is analyzed with an evolving model, consistent with an overflexible population description at the present signal-to-noise. The sign and magnitude of the shift can, however, depend on detector sensitivity and redshift reach as the population features become increasingly constrained directly by the data, motivating targeted diagnostics for future catalogs.
\end{abstract}

\maketitle

\section{Introduction}
Gravitational-wave (GW) observations of compact-binary coalescences provide an
independent route to cosmology through the \emph{standard siren} method~\cite{Schutz:1986gp,Holz:2005df}:
the strain amplitude directly encodes an absolute luminosity distance, $d_L$, inferred from the waveform with the binary’s intrinsic parameters.
Combining distances with redshift information allows one to constrain the
distance-redshift relation and therefore the Hubble constant $H_0$ and, more
generally, cosmological expansion and modified-propagation effects~\cite{Belgacem:2017ihm,Belgacem:2018lbp,LISACosmologyWorkingGroup:2019mwx}.
This approach is complementary to electromagnetic distance ladders~\cite{Riess:2021jrx} and to early Universe probes~\cite{Planck:2018vyg}, whose determination of $H_0$ are currently in strong tension,
and it is particularly valuable because it relies on different systematics.

Most sources in the latest GW catalog released by the LIGO-Virgo-KAGRA (LVK) collaboration~\cite{LIGOScientific:2014pky,VIRGO:2014yos,KAGRA:2020tym}, GWTC-4.0~\cite{LIGOScientific:2025hdt,LIGOScientific:2025slb}, are \emph{dark} standard sirens, as their individual host galaxies are not
identified and the redshift cannot be directly measured. In that case, redshift information must be obtained statistically.
\footnote{The only exception is the binary neutron star (BNS) coalescence GW170817~\cite{TheLIGOScientific:2017qsa,LIGOScientific:2017zic}, for which a coincident electromagnetic emission could be identified, leading to a direct measurement of $H_0$.}
Most of the cosmological information currently comes
from \emph{spectral siren} cosmology, in which
the distribution of source-frame masses and redshifts is used to
self-calibrate the redshift~\citep{Taylor:2011fs,Farr:2019twy,Mastrogiovanni:2021wsd,Ezquiaga:2022zkx}.
The basic mechanism is that the detector measures redshifted (detector-frame)
masses; converting to source-frame masses depends on $(1+z)$ and therefore on
cosmology. Distinct features in the mass spectrum---such as peaks or drops in the rate---act as ``anchors'' that correlate redshift with the
observed spectrum, enabling cosmological inference without host identification.

In particular, the most recent LVK analysis using the full GWTC-4.0 catalog relies largely on this principle~\cite{LIGOScientific:2025jau}.
In those analyses, evidence is found for the mass distribution of binary black holes (BBHs) to exhibit multiple features,
including pronounced structure beyond a single power law (also evident in the related population analysis~\cite{LIGOScientific:2025pvj}), which is precisely
what gives spectral sirens their leverage. At the same time, the reliance on
population features makes the inference sensitive to assumptions about the BBH
population model~\cite{Pierra:2023deu,Gennari:2025nho,LIGOScientific:2025jau}.

Indeed, among the most discussed sources of systematic uncertainty for spectral-siren
cosmology is the assumed form of the BBH mass function.
LVK analyses~\cite{LIGOScientific:2021aug,LIGOScientific:2025jau}, and subsequent GWTC-4.0-based spectral-siren analyses~\cite{Tagliazucchi:2026gxn,Pierra:2026ffj,Bertheas:2026odj,Gennari:2026dfy,Li:2026amt} (see also~\cite{Mali:2024wpq} for a thorough analysis on previous catalogs) so far have adopted \emph{redshift-independent}
mass distributions, and have explored
the impact of adopting different functional forms which allow for more spectral features.
An important complementary possibility, frequently raised in the literature~\cite{Mukherjee:2021rtw,Ray:2023upk,Heinzel:2024hva,Lalleman:2025xcs,Gennari:2025nho,Tenorio:2025nyt,Afroz:2025xpp}, is
that the BBH mass spectrum itself evolves with redshift.
The observational status of BBH mass evolution and the impact on cosmology is currently open:
analyses that fix cosmology generally find limited evidence for strong evolution in current
catalogs \cite{Ray:2023upk,Lalleman:2025xcs,LIGOScientific:2025pvj,Tenorio:2025nyt,Afroz:2025xpp},
while other works emphasize that even modest evolution can bias $H_0$ when the
mass model is used as the primary source of redshift information
\cite{Mukherjee:2021rtw,Agarwal:2024hld}.
This issue seems especially prominent in recent end-to-end mock-data studies that
find population mismodeling (specifically related to neglecting redshift evolution) to potentially
propagate into cosmological posteriors \cite{Agarwal:2024hld,Karathanasis:2026ldq}.
In particular, even when population analyses at fixed cosmology report no evidence for redshift evolution of the BBH mass spectrum, such evolution might remain effectively hidden while still biasing spectral-siren $H_0$ inference if not explicitly modeled~\cite{Karathanasis:2026ldq}. These conclusions, however, depend on the assumed simulation setting.

Motivated by these considerations, in this work we revisit GWTC-4.0 spectral-siren
constraints on $H_0$ while explicitly allowing for redshift evolution of the
main scales of the BBH mass spectrum. We adopt a parametric primary-mass model
matching the default LVK parametric choice~\cite{LIGOScientific:2025pvj}, namely one that includes a broken power law plus two Gaussian peaks, and we allow the
corresponding hyperparameters (peak locations and widths, power-law slopes and breaking point, and mixing fractions)
to evolve smoothly with redshift through flexible transition functions, inspired by Ref.~\cite{Lalleman:2025xcs}.
First, we study the possibility of evolution assuming a fixed Planck cosmology~\cite{Planck:2018vyg}.
We then quantify how this additional freedom affects the inferred $H_0$
relative to the redshift-independent baseline, and we present a set of diagnostics designed to make the effect of the
evolution assumptions transparent.

We compare the impact of redshift evolution with that of alternative modeling choices explored in recent GWTC-4.0 spectral-siren analyses, which introduce additional mass-spectrum features while keeping the population redshift-independent, in order to place the associated shifts in $H_0$ in a broader modeling context.

We finally compare with an injection-recovery study to clarify the origin of potential shifts in the inferred cosmological parameters and to explore their dependence on detector sensitivity and redshift reach.

\section{Methods and data}\label{sec:methods}

\subsection{Population model}
\label{sec:population_model}

To describe the BBH population, we adopt a parametric
probability density for source-frame component masses and redshift,
$p(m_1,m_2,z\mid\Lambda)$. This function encodes astrophysical assumptions
about masses and merger rate, and it is the central
ingredient carrying redshift information in spectral siren cosmology.

We model the mass distribution of BBHs using a joint probability
density in component masses and redshift,
\begin{equation}
p(m_1, m_2, z | \Lambda)
= p(z | \Lambda_z)\,
  p(m_1 | z, \Lambda_{m_1})\,
  p(m_2 | m_1, \Lambda_{m_2}),
\end{equation}

where $\Lambda$ denotes the full set of hyperparameters, partitioned into
redshift ($\Lambda_z$), primary-mass ($\Lambda_{m_1}$), and secondary-mass
($\Lambda_{m_2}$) parameters. The conditional mass distributions are normalized
at each redshift.

\vspace{0.5cm}
\paragraph{Primary-mass distribution.}
At fixed redshift, the primary black-hole mass $m_1$ is described by a
mixture of a broken power law and two Gaussian components, following the parametric model \textsc{Broken Power Law + 2 Peaks} adopted by the LVK population analysis~\cite{LIGOScientific:2025pvj} (hereafter \textsc{BPL+2P}),
\begin{equation}
\begin{split}
p(m_1 \mid z, \Lambda_{m_1})
\propto
& \, T(m_1) \,\Big[
\lambda_0(z)\,p_{\mathrm{BPL}}(m_1 \mid z)
+ \\
& \lambda_1(z)  p_{\mathrm{G},1}(m_1 \mid z)
+ \lambda_2(z)\,p_{\mathrm{G},2}(m_1 \mid z)
\Big]\,
,
\end{split}
\label{eq:mixture_pm1}
\end{equation}
with $\lambda_i(z)\ge 0$ and $\sum_i\lambda_i(z)=1$ by construction. The factor
$T(m_1)$ applies a smooth low-mass taper that suppresses the mass distribution at the lower edge $m_{1, \min}$ over a scale $\delta m_1$. The mixture is
evaluated within a fixed support $m_1\in[m_{1,\min},m_{1,\max}]$.
The broken-power-law component $p_{\mathrm{BPL}}$ has slopes $\alpha_1(z)$ and $\alpha_2(z)$ around
a break mass $m_b(z)$.
The Gaussian components $p_{\mathrm{G},i}$ ($i=1,2$) have redshift--dependent
means $\mu_i(z)$ and standard deviations $\sigma_i(z)$ and are truncated and
renormalized over the same interval $[m_{1,\min},m_{1,\max}]$.

\vspace{0.5cm}
\paragraph{Redshift evolution of mass hyperparameters.}
To allow the shape of the black-hole mass spectrum to vary with cosmic time,
we let selected primary-mass hyperparameters evolve smoothly with redshift,
following Ref.~\cite{Lalleman:2025xcs}.
For a generic hyperparameter $\phi$ we adopt
\begin{equation}
\phi(z) = \phi_0 + (\phi_\infty-\phi_0)\,S\!\left(\frac{z-z_\phi}{\Delta z_\phi}\right),
\;
S(x)=\frac{1}{2}\left[1+\tanh(x)\right].
\label{eq:sigmoid_evolution}
\end{equation}

Here $\phi_0$ and $\phi_\infty$ represent the asymptotic values of the parameter in the local Universe (under the condition $\Delta z_\phi \ll 1$) and at high redshift, while $z_\phi$ sets the characteristic redshift where the population transitions between these regimes and $\Delta z_\phi$ controls how gradual the transition is.
In this work, we evolve
\begin{equation}
\phi \in \{\alpha_1,\alpha_2, m_b, \mu_1,\sigma_1,\mu_2,\sigma_2\},
\end{equation}
each with its own $(z_\phi,\Delta z_\phi)$, and the mixture weights under the constraint $\lambda_i(z)\ge 0$ and $\sum_i\lambda_i(z)=1$ for all $z$. We enforce this by evolving the weights between two simplex end points, representing the mixture composition at low and high redshift. 
Specifically, we sample two simplex end points,
$\boldsymbol{\lambda}_0 \equiv \boldsymbol{\lambda}(z\simeq 0)$ and
$\boldsymbol{\lambda}_\infty \equiv \boldsymbol{\lambda}(z\to\infty)$, and
interpolate between them using a shared transition,
\begin{equation}
\boldsymbol{\lambda}(z)
=
\left[1-S\!\left(\frac{z-z_\lambda}{\Delta z_\lambda}\right)\right]\boldsymbol{\lambda}_0
+
S\!\left(\frac{z-z_\lambda}{\Delta z_\lambda}\right)\boldsymbol{\lambda}_\infty.
\end{equation}
Since both end points lie on the simplex, the weights remain on the simplex at
all redshifts.

Note that we allow redshift evolution only in the parameters controlling the main identifiable features of the primary-mass spectrum. This choice is motivated by the fact that this currently contains the most robustly identified structure, whereas the secondary-mass distribution is less well constrained and shows no comparably established features. It is therefore natural to first ask whether the spectral features that are already robustly inferred evolve with redshift, before introducing evolution in less constrained aspects of the joint mass distribution.

We also keep the mass-support parameters fixed. Parameters such as $m_{\min}$ and $m_{\max}$ live at the boundaries of the distribution and are constrained only by the subset of events close to those edges. At the current catalog size, inference in these regions is more sensitive to parametric assumptions, Monte-Carlo variance, and broad uncertainties than the inference of the main spectral features. Allowing these boundary parameters to evolve would therefore introduce additional weakly constrained degrees of freedom. 
Although support parameters can in principle correlate with cosmology, at current sensitivity the dominant spectral-siren information is carried by the well-populated features of the primary-mass spectrum rather than by the extreme tails of the distribution.
Nevertheless, evolution of the secondary-mass distribution and of support parameters, in particular $m_{\max}$, remains a well-motivated extension for future studies with larger catalogs.

\vspace{0.5cm}
\paragraph{Other population ingredients.}
The remaining components of the population model follow standard prescriptions.

The secondary mass $m_2$ is assumed redshift--independent and modeled with a power-law distribution truncated
at $m_2\le m_1$ and $m_2 > m_{2, {\rm min}}$, together with smooth tapering at the low-mass boundary. This treatment follows the default LVK
population models \cite{LIGOScientific:2025pvj}.

The merger-rate density as a function of redshift is described using the
Madau-Dickinson parametrization \cite{Madau:2014bja}, which captures the
rise and decline of the cosmic star-formation history through a broken
power-law form with a characteristic peak redshift.

We do not include an explicit population model of the black-hole spins, which corresponds to assuming that their distribution is uniform in magnitude and isotropic in orientation, as implied by the parameter estimation prior.\footnote{However, we explicitly checked that modeling spins with the using the \emph{Gaussian component spins} prescription adopted in the GWTC-4.0 population analysis~\cite{LIGOScientific:2025pvj} has no impact on the result.}

Full mathematical details of these ingredients are provided in
Appendix~\ref{app:population_model}.

\vspace{0.5cm}
\paragraph{Summary of hyperparameters.}
The full set of hyperparameters is denoted by $\Lambda$ and grouped as
\[
\Lambda=\{\Lambda_{\rm c},\Lambda_{z},\Lambda_{m}\},
\]
describing cosmology,
merger-rate evolution, and the mass spectrum, respectively.
Throughout we assume a flat $\Lambda$CDM cosmology with two free parameters,
$\Lambda_{\rm c}=\{H_0,\Omega_m\}$.
The merger-rate density is controlled by the three parameters
$\Lambda_{z}=\{\gamma,\kappa,z_p\}$ (see below).
For the mass model, the low-redshift end points comprise
\begin{equation}\label{eq:zero_endpoints}
\begin{split}
\Lambda_{m,0}=&\, \{\alpha_{1,0},\alpha_{2,0},m_{b,0},\mu_{1,0},\sigma_{1,0},\mu_{2,0},\sigma_{2,0},
m_{1,\min},m_{1,\max},\\
& \delta m_1,\beta,m_{2,\min},\delta m_2,\boldsymbol{\lambda}_0\},
\end{split}
\end{equation}
where $\boldsymbol{\lambda}_0=(\lambda_{0,0},\lambda_{1,0},\lambda_{2,0})$ are the
mixture weights at $z\simeq 0$.

We also compare results with those obtained with a nonevolving mass spectrum; in this case, the mass hyperparameters are restricted to the low-redshift end points in Eq.~\ref{eq:zero_endpoints}. 

Redshift evolution is introduced for
$\phi\in\{\alpha_1,\alpha_2, m_b, \mu_1,\sigma_1,\mu_2,\sigma_2\}$ via
Eq.~(\ref{eq:sigmoid_evolution}), each with its own triplet
$(\phi_\infty,z_\phi,\Delta z_\phi)$ . 
The mixture weights evolve between end point simplices
$\boldsymbol{\lambda}_0$ and $\boldsymbol{\lambda}_\infty$ using a shared
transition $(z_\lambda,\Delta z_\lambda)$, with
$\boldsymbol{\lambda}_\infty=(\lambda_{0,\infty},\lambda_{1,\infty},\lambda_{2,\infty})$.

In total, the model has
$N_\Lambda=2\;(\mathrm{cosmology})+3\;(\mathrm{rate})
+15\;(\mathrm{mass}\;z\simeq0)+21\;(\mathrm{shape\;evolution})+4\;(\mathrm{weight\;evolution})
=45$
free hyperparameters, where the mass low-$z$ count includes
$m_{1,\min}$ and $m_{2,\min}$ (sampled through auxiliary variables as described
in Appendix~\ref{app:priors}) and the simplex end points each contribute two
independent degrees of freedom.

\subsection{Prior choices}\label{sec:priors}

We adopt weakly informative priors for all population and cosmological
hyperparameters, summarized in Appendix~\ref{app:priors}. 

For variables with hard uniform priors defined in a given interval, rather than sampling such quantities directly, we sample an
unconstrained ``raw'' variable and map it smoothly to the target interval using a logistic transform. 
This removes hard boundaries in the
sampling space, improving Hamiltonian Monte Carlo geometry and numerical stability.
We use the same strategy for cosmological parameters and
for bounded population hyperparameters. 
For some parameters we apply the same construction in log space.
Details are given in Appendix~\ref{app:priors}.

For the nonredshift-evolving case, the prior for the mass hyperparameters coincides with those chosen for the low-redshift end points.

For redshift evolution, we parametrize the high-redshift endpoint through an
additive difference $\Delta\phi \equiv \phi_\infty-\phi$, and place
zero-mean normal priors on $\Delta\phi$ with scale
$\sigma_{\Delta\phi}$. Transition redshifts $z_\phi$ are assigned uniform
priors, while transition widths $\Delta z_\phi$ are sampled in log space and
then exponentiated, producing a positive prior truncated to a finite
interval. 
Our default priors allow the transition to occur over $z\in [0.05$–$1.5]$, sufficient to cover the redshift span of the data, with widths $\Delta z_\phi\in[0.05,2]$, corresponding to either relatively sharp or extended evolution across cosmic time.

For the mixture weights we sample both end point vectors $\boldsymbol{\lambda}_0$ and $\boldsymbol{\lambda}_\infty$ from Dirichlet distributions. 

\subsection{Statistical framework and sampling details}\label{sec:stats}

We infer the population and cosmological hyperparameters $\Lambda$
using a hierarchical Bayesian framework
\cite{Mandel:2018mve,Thrane:2018qnx,Vitale:2020aaz}. 

The data for each event consist of the LVK
posterior samples obtained from single-event parameter estimation
using a reference prior.
The population likelihood combines the information from all detected
events and accounts for selection effects.
For each event, the likelihood contribution is computed by
reweighting the posterior samples from detector-frame parameters
to the source-frame population model. This requires mapping
detector-frame quantities (detector-frame masses and luminosity
distance) to source-frame parameters using the cosmology assumed
in the population model; in presence of features in the source-frame mass distribution, this mapping provides constraining power on cosmological parameters, effectively breaking the mass-redshift degeneracy of the GW waveform. 

Selection effects are incorporated through the detection efficiency,
defined as the fraction of sources drawn from the population model
that would be detected by the search pipeline. We estimate this
quantity using Monte Carlo integration over simulated injections,
reweighting each injection according to the population model and
the injection distribution~\cite{Tiwari:2017ndi,Farr:2019rap}.

Following Refs.~\cite{Talbot:2023pex,Essick:2022ojx,Heinzel:2025ogf}, we monitor the
Monte Carlo variance of the likelihood estimator and impose the
standard requirement that the resulting variance of the log-likelihood
remains below unity to avoid biases in population inference.

Posterior sampling of the hyperparameters is performed with an extension of the \texttt{python} package \pymcpop~\cite{pymcpop}   introduced by Ref.~\cite{Mancarella:2025uat}. 
The package employs the no-u-turn sampler (NUTS)~\cite{JMLR:v15:hoffman14a,Brooks_2011} with GPU acceleration leveraging either \texttt{PyMC}~\cite{pymc2023} or \texttt{numpyro}~\cite{phan2019composable,bingham2019pyro}. We employ the \texttt{numpyro} version for production runs.

We run four independent chains of 1,000 steps each, after discarding a burning phase of 1,000 additional steps, and verify convergence by checking the Gelman-Rubin $\hat{r}$ statistics~\cite{10.1214/ss/1177011136}, ensuring that $\hat{r}\approx1.0$ for all variables.

Full mathematical details of the likelihood
construction and evaluation are provided in Appendix~\ref{app:stats}.

\section{Results}\label{sec:results}

\subsection{Data}\label{sec:data}

We analyze the subset of GWTC-4.0 events used in the LVK population analysis~\cite{LIGOScientific:2025pvj},
restricted to binary black holes (BBHs) with false-alarm rate $\mathrm{FAR}<1\,\mathrm{yr}^{-1}$.
This yields a total of $N_{\rm obs}=153$ BBH detections spanning the LVK observing
runs considered in the GWTC-4.0 population release. Specifically, 10 events are from the O1-O2 observing runs~\cite{LIGOScientific:2018mvr}, 36 from the O3a run~\cite{LIGOScientific:2020ibl}, 23 from O3b~\cite{KAGRA:2021vkt}, 84 from O4a~\cite{LIGOScientific:2025slb}.

We use the LVK-provided posterior samples from the event-level
parameter-estimation (PE) analyses to evaluate the likelihood as described in Sec.~\ref{sec:stats} and Appendix~\ref{app:stats}.
These are available at~\cite{ligo_scientific_collaboration_and_virgo_2025_17014085, ligo_scientific_collaboration_and_virgo_2023_8177023, ligo_scientific_collaboration_and_virgo_2022_6513631}.
We use samples obtained with the \texttt{IMRPhenomXPHM} waveform model for events in O1 to O3, and \texttt{IMRPhenomXPHM\_SpinTaylor} for O4a events.

For selection effects, we compute the detection efficiency by reweighting detected
injections as described in Sec.~\ref{sec:stats} and Appendix~\ref{app:stats}. We use the publicly released injection sets accompanying the GWTC-4.0 population
products, available at~\cite{ligo_scientific_collaboration_2025_16740128} and described in~\cite{Essick:2025zed}; see also~\cite{Essick:2023toz} for additional details.

\subsection{Result at fixed cosmology}

\begin{figure*}[t]
\centering
\includegraphics[width=0.49\textwidth]{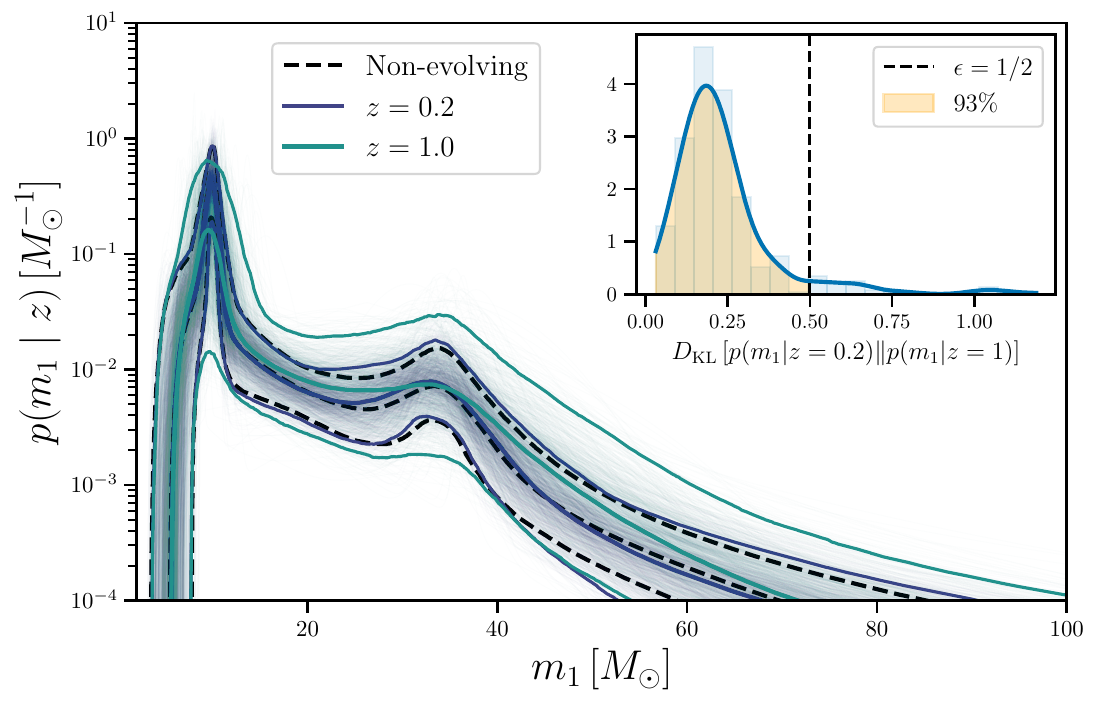}
\hfill
\includegraphics[width=0.49\textwidth]{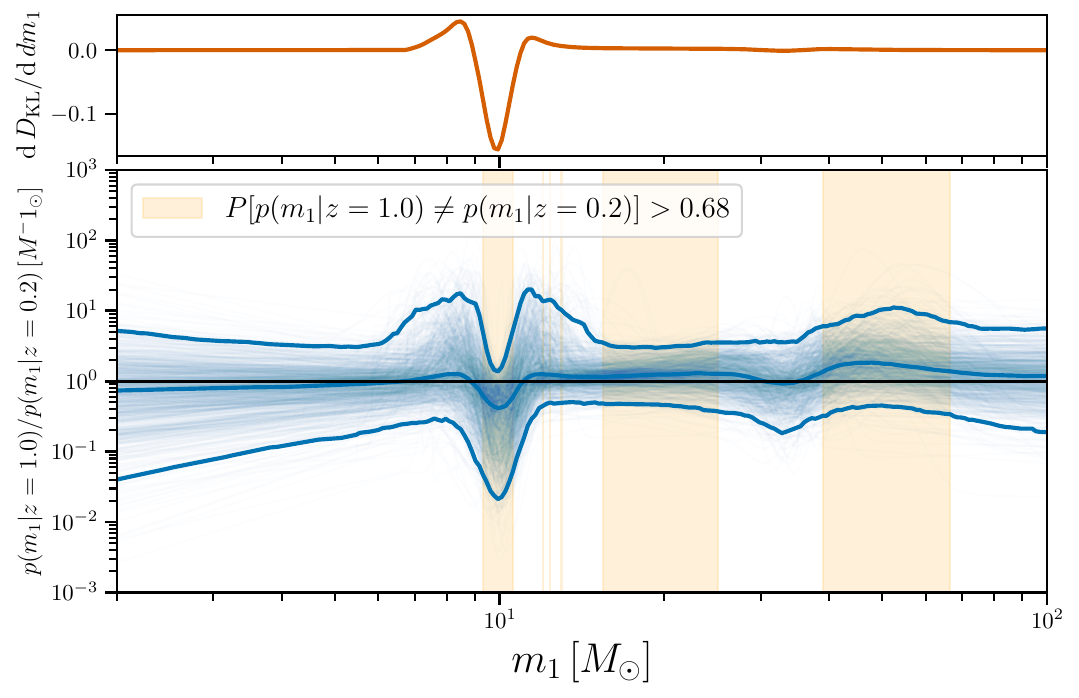}
\caption{Left: marginal posterior $p(m1|z)$ with bands at two different redshift slices.
The thin colored lines represent individual 1D marginal distribution draws from the hyper-posterior. The thick lines indicate the 5\,$\%$, 50\,$\%$ and 95\,$\%$ percentiles. The black dashed lines correspond to analogous results obtained using a BPL-2P nonevolving mass model. The inset is showing the KL histogram between $p(m1|z=0.1)$ and $p(m1|z=1)$ draws and the fraction that satisfies our chosen TVD bound represented by the yellow region and the black dotted vertical line. Right: Posterior distribution of the ratio between $p(m1|z)$ at two different redshift slices. The bottom panel shows individual realizations with the 5\,$\%$, 50\,$\%$ and 95\,$\%$ percentiles. The yellow regions indicates the $m_1$ support for which the 68\,$\%$ of this ratio draws differ from one. The top panel represents the derivative of the median KL divergence between the two different $p(m1|z)$ with respect to $m_1$.}
\label{fig:pm1_marginal_at2_redshifts}
\end{figure*}

We start with an analysis where cosmology is fixed to Planck values~\cite{Planck:2018vyg} to assess the presence of any redshift evolution.

Figure~\ref{fig:pm1_marginal_at2_redshifts} shows the reconstructed marginal distribution $p(m1 | z, \Lambda)$ at $z=0.2$ (purple) and $z=1$ (cyan), while the result for the nonevolving case is also shown with dashed black lines. 
The low-redshift slice matches closely the nonevolving case, while the slice at $z=1$ shows no visual evidence of a clear displacement of the peaks, but an overall broadening of the constraint, especially above the $\sim 35 M_{\odot}$ peak. We make these considerations more quantitative below.

\vspace{0.5em}
\paragraph{Posterior Predictive Checks.}
\label{sec:ppc}
First, we assess the adequacy of the population model by performing posterior predictive checks (PPCs)~\cite{Fishbach:2019ckx,Romero-Shaw:2022ctb,Miller:2026buq}. %
These evaluate whether data generated from the fitted model resemble the observed catalog.
For each posterior draw of hyperparameters $\Lambda_k$, we construct synthetic catalogs by reweighting the single-event posterior samples to obtain an observedlike catalog, and by resampling the injection set according to the population model to generate predicted catalogs.

We then compare these catalogs using a discrepancy statistic $T$, specifically the Kolmogorov--Smirnov (KS) distance between the cumulative distribution functions of the observedlike ($y_{\rm obs}^{(k)}$) and predicted ($y_{\rm rep}^{(k)}$) catalogs for the primary source-frame mass $m_1$ or the redshift $z$.

For each posterior draw $\Lambda_k$, this procedure yields two discrepancy measures,
\begin{align}
D_{\rm obs-rep}^{(k)} &= T\left(y_{\rm obs}^{(k)}, y_{\rm rep}^{(k)}\right)\, , \, D_{\rm rep-rep}^{(k)} = T\left(y_{\rm rep}^{(k)}, y_{\rm rep'}^{(k)}\right).
\end{align}

The first quantity, $D_{\rm obs-rep}^{(k)}$, measures the difference between the observedlike catalog and a model-predicted catalog.
However, even if the model is correct, this discrepancy does not vanish, as both catalogs are finite Monte Carlo realizations and therefore differ due to sampling fluctuations.
The second quantity, $D_{\rm rep-rep}^{(k)}$, provides the natural reference scale for discrepancies expected under the model itself.
If the model is correct, the observed catalog should behave statistically like another realization drawn from the predictive distribution, implying
$D_{\rm obs-rep}^{(k)} \sim D_{\rm rep-rep}^{(k)}$.
Conversely, if the model is misspecified, the observed catalog will tend to differ more strongly from model-generated catalogs. 

We therefore define the posterior predictive $p$-value as
\begin{equation}
p_{\rm ppc} =
P\left(D_{\rm rep-rep} \ge D_{\rm obs-rep}\right),
\end{equation}
that is, the fraction of posterior draws for which the discrepancy between two replicated catalogs exceeds that between the observedlike and replicated catalogs.
Values of $p_{\rm ppc}$ near $0.5$ indicate that the observed catalog is statistically typical relative to catalogs generated under the model, whereas values close to $0$ or $1$ indicate that it is atypical.
Following Ref.~\cite{Miller:2026buq}, we consider the symmetrized posterior predictive $p$-value
\begin{equation}
\bar p_{\rm ppc} = 1 - 2\,\left| p_{\rm ppc} - 0.5\right| \, ,
\end{equation}
and adopt $\bar p_{\rm ppc} > 0.05$ as a criterion for no evidence of discrepancy.
Under this condition, the model can be considered consistent with the data according to the chosen statistic $T$. As discussed in~\cite{meng1994posterior}, $ \bar p_{\rm ppc}$ satisfies
\begin{equation}\label{eq:Pppc_def}
P(\bar p_{\rm ppc} \leq \alpha) \leq \alpha \,.
\end{equation}
Therefore, if $\bar p_{\rm ppc} \leq 0.05$, the model may be rejected with a type~I error probability not exceeding $\alpha = 0.05$.\footnote{Note that the $p$-value provides a necessary but not sufficient condition for model adequacy; in particular, if the condition in Eq.~\ref{eq:Pppc_def} is satisfied, we can only conclude that there is no evidence of a discrepancy given the data and models under consideration.}

For the global distributions, we obtain posterior predictive $p$-values
$\bar p_{\rm ppc}(m_1) \approx 0.96$ ($\bar p_{\rm ppc}(m_1) \approx 0.96$) and
$\bar p_{\rm ppc}(z) \approx 0.76$ ($\bar p_{\rm ppc}(z) \approx 0.68$)
for the redshift-evolving (nonredshift-evolving) model.
These results indicate that the observed catalog is consistent with typical realizations of both models.
We further test the evolving model by repeating the PPC in three redshift bins.
The resulting symmetrized posterior predictive $p$-values are $0.96$, $0.80$, and $1.00$ for the bins $z<0.5$, $0.5\le z<1$, and $z\ge1$, respectively.

Overall, these posterior predictive checks show that both the nonevolving and evolving population models provide statistically adequate descriptions of the current catalog within the sensitivity of the data.
We provide plots of the cumulative distributions in Appendix~\ref{app:ppc}.

\begin{figure*}[th]
\centering
\includegraphics[width=0.24\linewidth]{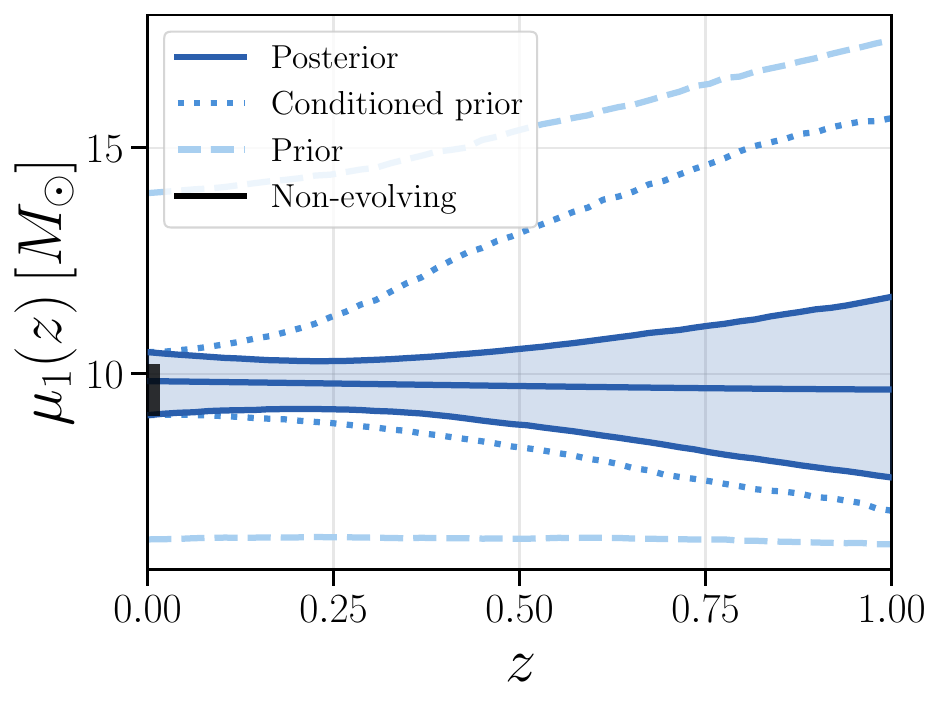}
\hfill
\includegraphics[width=0.24\linewidth]{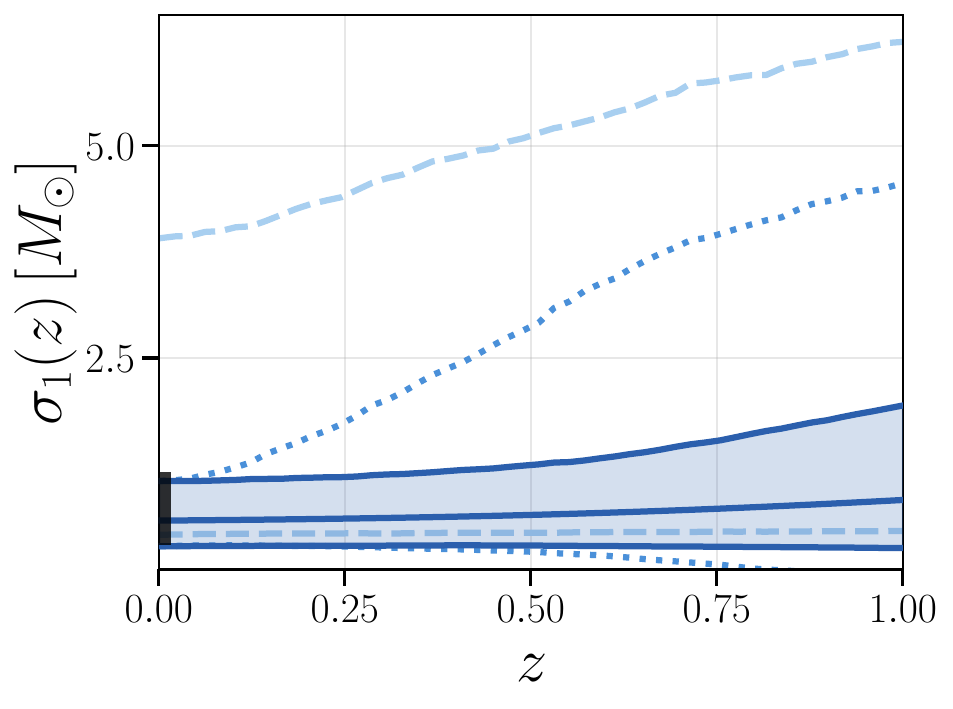}
\hfill
\includegraphics[width=0.24\linewidth]{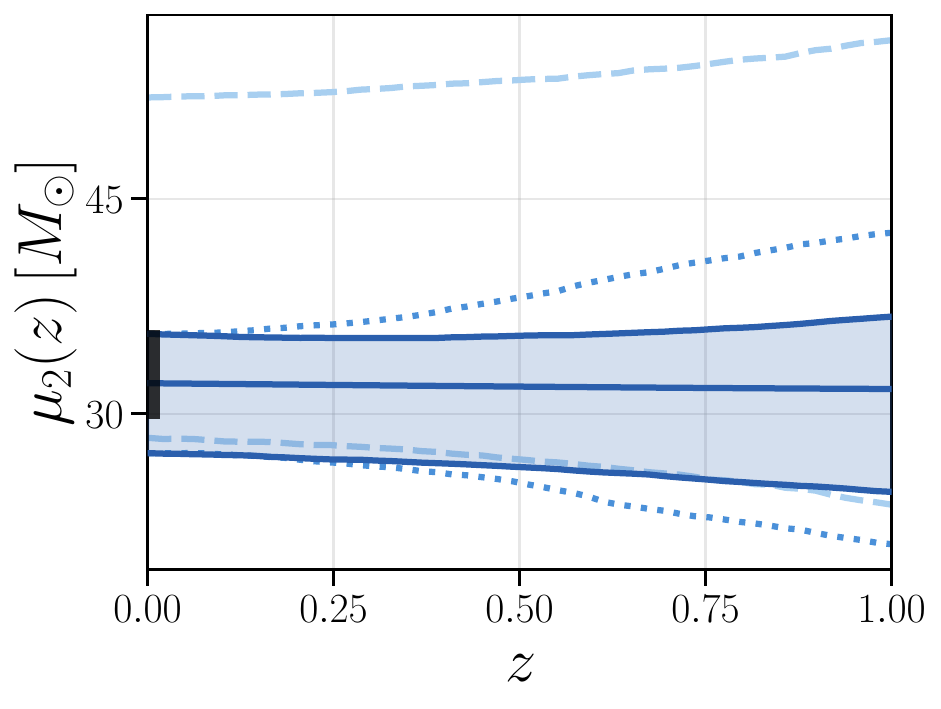}
\hfill
\includegraphics[width=0.24\linewidth]{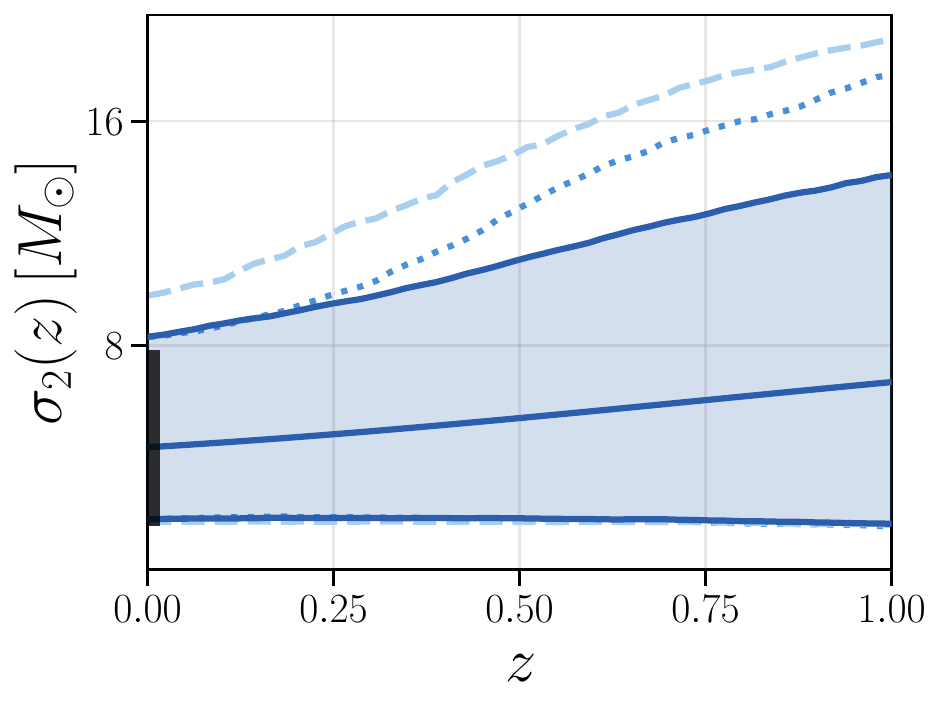}
\caption{Inferred values of $\mu_{1/2}(z)$ and $\sigma_{1/2}(z)$. The thick lines indicate the median and 95\,$\%$ credible bounds of the posteriors. The dashed and dotted lines correspond, respectively, to the 95\,$\%$ credible bounds of the priors and priors conditioned on the inferred posteriors at $z=0$. The black vertical bands at $z=0$ represent the 90\,$\%$ credible levels of nonevolving model posteriors.}
\label{fig:param_evo}
\end{figure*}

\vspace{0.5em}
\paragraph{Overall evolution of the BBHs primary mass spectrum.}
\label{sec:KL_overall}

Next, we quantitatively assess the absence of evidence for an overall evolution of the mass spectrum.
We first investigate possible redshift evolution by comparing the one-dimensional marginal posterior distributions $p(m_1 \mid z)$ at two redshift slices, $z = 0.2$ and $z = 1$.

A way to quantify the difference between two probability distributions $p$ and $q$ defined on the same parameter space $\Theta$ is through the supremum (or $L_\infty$) distance:
\begin{equation}
\label{eq:TVD}
\delta(p, q) = \sup_{\theta \in \Theta} \, \lvert p(\theta) - q(\theta) \rvert \, ,
\end{equation}
which represents the largest absolute pointwise difference between the two distributions.
This quantity can be related to more standard metrics such as the Kullback–Leibler (KL) divergence (that also have the advantage of being more computationally tractable in arbitrary dimensions):
\begin{equation}
\label{eq:KL}
D_{\mathrm{KL}}(p \parallel q)
=\int_{\Theta}
p(\theta) \,
\log \frac{p(\theta)}{q(\theta)}\,\mathrm{d}\theta \, ,
\end{equation}
through Pinsker’s inequality:
\begin{equation}
\label{eq:Pinsker_bound}
\delta(p, q) \leq \sqrt{\dfrac{1}{2}D_{\mathrm{KL}}(p \parallel q)} \, ,
\end{equation}
which provides an upper bound on the supremum distance in terms of the KL divergence.
By introducing a threshold $\epsilon$ such that $D_{\mathrm{KL}} \leq \epsilon$, one can bound the KL divergence and, consequently, the maximum absolute difference between $p$ and $q$.
Our choice of $\epsilon$ is inspired by Markov chain mixing arguments~\cite{Levin, Montenegro, Aldous}. In particular, we set $\sqrt{\epsilon/2} = 1/2$, corresponding to a commonly used criterion for the total variation mixing time, i.e., the scale at which a Markov chain is considered to be within a “small distance” from stationarity. In this sense, the condition $\epsilon \leq 1/2$ defines a conservative threshold below which two distributions can be regarded as effectively indistinguishable at the level of typical sampling fluctuations.

We construct a distribution of $D_{KL}$ between $p(m_1 \mid z)$ at $z = 0.2$ and $z = 1$ as follows.
First, we divide each ensemble of sampled curves at fixed redshift into ordered “stripes”.\footnote{Bounding the KL divergence for every possible combination of $p(m_1 \mid z=0.2)$ and $p(m_1 \mid z=1)$ draws from the hyper-posterior can be overly restrictive. In regimes with large uncertainty in $p(m_1 \mid z)$, the results can be sensitive to outliers, leading to artificially large KL values due to random mismatches between individual realizations, even in the absence of genuine redshift evolution.} These stripes are constructed by ranking individual realizations according to their $L_2$ distance from the median curve, and then partitioning the ordered ensemble into subsets of equal size.
For each stripe, we define a representative curve by averaging all realizations within the subset. This procedure yields a reduced set of smooth representative curves, suppressing stochastic fluctuations. We then compute the KL divergence between 400 corresponding ordered stripes of $p(m_1 \mid z)$ at $z=0.2$ and $z=1$. This choice ensures a sufficient number of draws ($\sim 10$) within each stripe to obtain a reliable representative curve.
The inset in Fig.~\ref{fig:pm1_marginal_at2_redshifts} shows the distribution of KL divergence values across the stripes. The vertical dashed line indicates the chosen threshold $\epsilon$, above which the distributions are considered discrepant.
We find that $\sim 93\,\%$ of the pairs of stripes are consistent with the nonevolving hypothesis.\footnote{The specific value obtained is of course sensitive to the binning choice. As a sanity check, we repeat the analysis in the extreme case of only four draws per stripe. The consistency decreases to ${\sim}75\,\%$, remaining well within expectations for no evolution.}

\vspace{0.5em}
\paragraph{Evolution at different mass scales}
\label{sec:KL_mass_scale}

In general, the KL divergence provides an integrated measure of the difference between two distributions over their full support. Therefore, on its own, it cannot probe differential evolution occurring at specific mass scales.
Inspired by Ref.~\cite{Tenorio:2025nyt}, we therefore consider the ratio $p(m_1 \mid z=1)/p(m_1 \mid z=0.2)$ and quantify how much it deviates from the nonevolving scenario as a function of $m_1$. This is shown in the right panel of Fig.~\ref{fig:pm1_marginal_at2_redshifts}, with the upper plot showing the derivative of the KL divergence as a function of mass, which is expected to vary more rapidly around evolving features.
The $95\,\%$ credible interval of the ratio is consistent with no evolution across the entire range of $m_1$, indicating that the nonevolving hypothesis is consistent at all masses.

However, we identify three mass ranges for which the $68\,\%$ credible interval of the ratio excludes the nonevolving expectation (shaded orange bands in the right panel of Fig.~\ref{fig:pm1_marginal_at2_redshifts}). These regions are largely determined by the two peaks of the mass spectrum (as also indicated by the behavior of the KL derivative in the upper panel).
Specifically, this behavior arises because, as the redshift increases, the widths of the Gaussian peaks broaden while their amplitudes decrease---this is caused by an overall normalization effect in the primary mass distribution rather than an effective redshift evolution of the mixture weights. At the same time, the positions of the peaks remain unchanged with redshift, consistent with previous studies~\cite{Lalleman:2025xcs}.
More observations will help determine whether such a trend becomes statistically significant; at the current catalog size, it is consistent with being driven by the increased prior volume of the evolving model.

In the context of a parametric model, however, it is not surprising that regions associated with explicitly parametrized features (the two Gaussian peaks) are more likely to be associated with a potential discrepancy.
More data-driven studies (e.g.,~\cite{Tenorio:2025nyt}), which do not rely on specific parametric forms, could provide more model-independent indications of tentative redshift evolution (note that Ref.~\cite{Tenorio:2025nyt} finds a potential evolution near the low-mass peak of the total mass spectrum).

\vspace{0.5em}
\paragraph{Evolution at single population parameter level}
\label{sec:param_evo}

Finally, following Ref.~\cite{Lalleman:2025xcs}, we examine the redshift evolution of specific hyperparameters. We focus on the position $\mu$ and standard deviation $\sigma$ of the two Gaussian components, which are the most relevant for cosmology and, as shown in Fig.~\ref{fig:pm1_marginal_at2_redshifts}, the parameters for which some (nonstatistically significant) discrepancies may appear.\footnote{Results for the other mass model parameters show similar trends.} 
The results are shown in Fig.~\ref{fig:param_evo}, where we display the redshift evolution of $\mu_{1,2}$ and $\sigma_{1,2}$ as a function of redshift (filled bands).
To explicitly test whether the absence of evolution is driven by the assumed parametric form of the redshift transition, we follow Ref.~\cite{Lalleman:2025xcs} and compare to the prior conditioned on the values at redshift $z=0$. We find that all hyperparameter posteriors are narrower than this induced prior, indicating that the model effectively constrains the absence of evolution across redshift, rather than merely extrapolating from the values at $z=0$ (dotted lines). This effect is particularly clear for the $\sim 10\,M_{\odot}$ peak, while the standard deviation of the $\sim 35\,M_{\odot}$ peak exhibits a less informative behavior.

In general, we find no conclusive evidence for redshift evolution in any of the mass hyperparameters.


\subsection{Results with free cosmology}

\begin{figure}
\centering
\includegraphics[width=\linewidth]{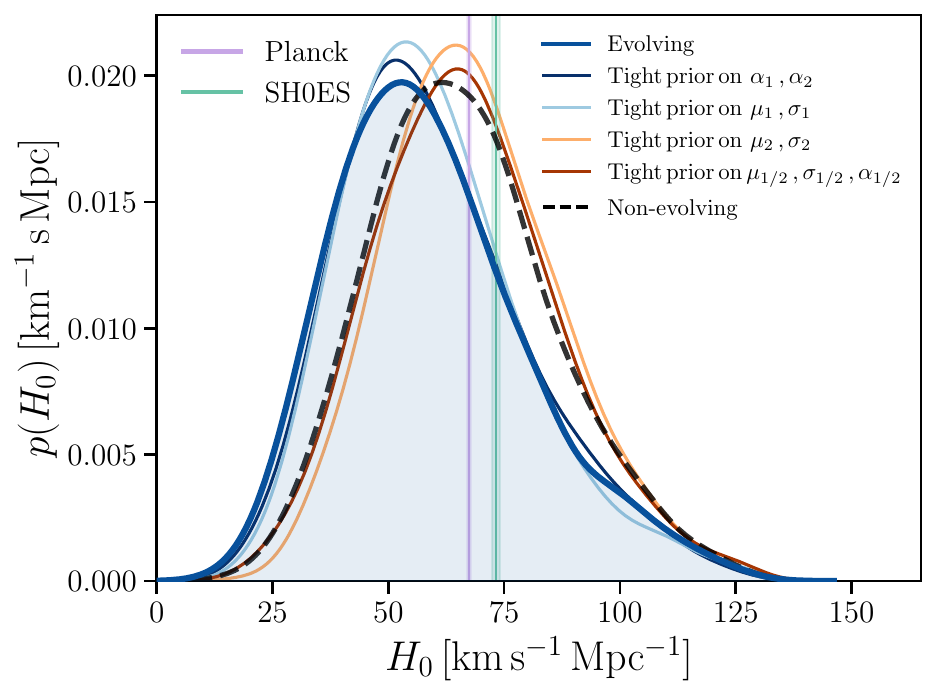}
\caption{One-dimensional $H_0$ posteriors. Different colors refer to different choice of the priors on evolution parameters. The blue curve represents our evolving BPL+2P result while the black dashed is related to the nonevolving case. Green and pink shaded bands respectively show local and early Universe 68\% constraints on $H_0$.
}\label{fig:H0marg}
\end{figure}

\begin{table}
\centering
\setlength{\tabcolsep}{3pt}
\renewcommand{\arraystretch}{2.}
\scriptsize
\begin{tabular}{l|c|c|c}
\toprule
Model & $H_0$ & $\Delta H_0 / \sigma_{H_0}$ & $|\Delta H_0|$\\
\midrule
\textbf{Evolving} \textsc{BPL+2P} & $56.4^{+23.2}_{-17.1}\ (56.4^{+42.6}_{-26.2})$ & $-0.30$ & $8.42$\\
\midrule
Tight prior on $\alpha_1$, $\alpha_2$ & $57.2^{+22.9}_{-16.5}\ (57.2^{+41.0}_{-25.3})$ & $-0.27$ & $7.60$\\
Tight prior on $\mu_1$, $\sigma_1$ & $57.3^{+21.0}_{-16.0}\ (57.3^{+39.0}_{-24.8})$ & $-0.28$ & $7.50$\\
Tight prior on $\mu_2$, $\sigma_2$ & $65.8^{+20.2}_{-18.4}\ (65.8^{+36.3}_{-28.7})$ & $+0.04$ & $0.98$\\
\makecell[l]{Tight prior on $\mu_{1/2}$, $\sigma_{1/2}$, \\ $\alpha_{1/2}$}
& $66.9^{+20.0}_{-16.9}\ (66.9^{+34.7}_{-26.8})$ & $+0.08$ & $2.13$\\
\midrule
Nonevolving \textsc{BPL+2P} & $64.8^{+21.1}_{-18.1}\ (64.7^{+37.2}_{-28.2})$ & -- & --\\
\bottomrule
\end{tabular}
\caption{$H_0$ Posterior statistics for all model configurations. Values are reported as median with 68\% and 90\% credible intervals in units of $\mathrm{km\,s^{-1}\,Mpc^{-1}}$. The corresponding normalized and absolute $H_0$ shifts are respectively defined as
$(H_0^{\rm evol}-H_0^{\rm non})/\sqrt{\left(\sigma_{H_0}^{\rm non}\right)^2+\left(\sigma_{H_0}^{\rm evol}\right)^2 }$ and $
|H_0^{\rm evol}-H_0^{\rm non}|$ in units of the statistical uncertainty $\sigma_{H_0}$ and $\mathrm{km\,s^{-1}\,Mpc^{-1}}$.
}
\label{tab:H0vals}
\end{table}

We next move to opening the parameter space to $\{H_0, \Omega_m\}$. 

We find that all the conclusions about the mass distribution are analog to the cosmology-fixed case, although with slightly broader uncertainties due to the additional freedom. 
In particular, the distribution of KL divergences between $p(m_1 \mid z)$ at $z=0.2$ and $z=1$ in ordered stripes, described in Sec.~\ref{sec:KL_overall}, yields a $94\%$ of realizations compatible with no evolution.
The relative figures can be found in Appendix~\ref{app:cosmo_evol}.

As for the Hubble constant, we obtain a value of $H_0=56.4^{+23.2}_{-17.1}$ $\mathrm{km\,s^{-1}\,Mpc^{-1}}$ (median and 68\% CL). 
For comparison, we also perform an analysis with the nonredshift-evolving model, finding $H_0=64.8^{+21.1}_{-18.1}$ $\mathrm{km\,s^{-1}\,Mpc^{-1}}$. 
The inferred $H_0$ posterior for the redshift-evolving case is slightly shifted toward lower values relative to the redshift-independent baseline, although this shift remains well below the statistical uncertainty.
Expressed in units of the combined uncertainty—computed by approximating each asymmetric 68\% credible interval by its average half-width and adding the uncertainties in quadrature—the shift is $\approx 0.3\,\sigma$, which shows excellent agreement.

\vspace{0.5em}
\paragraph{Comparison to varying spectral features at fixed redshift.}

We adopt the $\approx 0.3\,\sigma$ shift as a representative scale for the variation induced by redshift evolution, and compare it to shifts arising from alternative modeling assumptions in the literature, all based on introducing different redshift-independent spectral features (see Table~\ref{tab:H0_shifts}). For this purpose, we take as a reference the LVK spectral-siren analysis with the event selection and model closest to those references, namely their \textsc{Multi-Peak} model, which finds $H_0=77.1^{+40.8}_{-26.3}$ $\mathrm{km\,s^{-1}\,Mpc^{-1}}$~\cite{LIGOScientific:2025jau}.

In Ref.~\cite{Tagliazucchi:2026gxn}, which explores an extended range of spectral features based on B-splines, most inferred values lie within $\sim0.1$--$0.5\sigma$ of the LVK result, with the most discrepant configuration reaching $\sim0.7\sigma$. In terms of absolute shifts, the inferred values differ from the LVK reference by $\sim4$--$25\ \mathrm{km\,s^{-1}\,Mpc^{-1}}$, corresponding to roughly $\sim0.5$--$3$ times the shift induced by redshift evolution in our reference analysis, with most cases being around $\sim2$ times that scale.
Reference~\cite{Pierra:2026ffj} finds a value very close to the LVK reference ($\sim0.04\,\sigma$) when adding a third Gaussian feature around $\sim60\,M_{\odot}$.
In Ref.~\cite{Bertheas:2026odj}, which considers a broader set of cosmological and population models allowing for sharper mass features, the inferred values span a wider range, with shifts typically between $\sim0.3$ and $0.7\sigma$, and the most discrepant configuration reaching $\sim0.8\sigma$. In terms of absolute shifts, the inferred central values differ from the LVK reference by $\sim12$--$26\ \mathrm{km\,s^{-1}\,Mpc^{-1}}$, corresponding to roughly $\sim1.5$--$3$ times the shift induced by redshift evolution in our reference analysis.
Finally, Ref.~\cite{Gennari:2026dfy}, which investigates spline-based reconstructions and subpopulation models, reports shifts typically ranging from negligible ($<0.1\,\sigma$) up to $\sim0.8\,\sigma$, corresponding to absolute shifts of order a few to $\sim25\ \mathrm{km\,s^{-1}\,Mpc^{-1}}$, i.e., up to roughly $\sim3$ times the shift induced by redshift evolution in our reference analysis. However, the most extreme spline configurations explored in that work lead to much larger excursions, reaching shifts of up to $\sim2.5\,\sigma$.

More importantly, Ref.~\cite{Gennari:2026dfy} explicitly points out that allowing excessive flexibility in the reconstruction of the mass distribution can lead to extremely large shifts in the inferred value of $H_0$, together with population models that become difficult to interpret physically. In particular, when too many spline degrees of freedom are introduced, the cosmological parameters can effectively be used to compress the inferred population into a small number of sharp features, producing very large values of $H_0$.
Determining the appropriate amount of allowed variability in the mass distribution, as well as robust metrics to quantify and control model complexity, therefore remains an open problem. 

Despite the statistical insignificance of the shifts discussed above, this suggests that uncontrolled population-model flexibility may constitute a potentially larger source of systematic uncertainty for spectral-siren cosmology than a putative redshift evolution of the mass spectrum, at least at current sensitivity.
Moreover, visual inspection of the corresponding posteriors shows that different assumptions for the population model can substantially modify the overall shape of the inferred $H_0$ distribution---including its width, skewness, and tails---often much more strongly than the redshift evolution considered in this work.
Overall, the shift associated with redshift evolution in our modeling is smaller, and fully consistent with, the level of variation induced by alternative assumptions for redshift-independent mass features in recent spectral-siren analyses.

\begin{table}
\centering
\setlength{\tabcolsep}{3pt}
\renewcommand{\arraystretch}{2.}
\begin{tabular}{l|c|c}
\toprule
Model & $\Delta H_0 / \sigma_{H_0}$ & $|\Delta H_0|$\\
\midrule
B-spline spectral features~\cite{Tagliazucchi:2026gxn} & $0.05$--$0.7$ & $4$--$25$\\
Third Gaussian peak at ${\sim}60\,M_\odot$~\cite{Pierra:2026ffj} & ${\sim}0.04$ & ${\sim}2$\\
Sharper mass features~\cite{Bertheas:2026odj} & $0.3$--$0.8$ & $12$--$26$\\
Spline \& subpopulation models~\cite{Gennari:2026dfy} & $<0.1$--$0.8$ & few--$25$\\
\bottomrule
\end{tabular}
\caption{Summary of relative and absolute $H_0$ shifts induced by different (redshift independent) modeling assumptions explored in the literature, respectively expressed in units of the statistical uncertainty $\sigma_{H_0}$ and $\mathrm{km\,s^{-1}\,Mpc^{-1}}$. All comparisons are relative to the LVK \textsc{Multi-Peak} spectral siren result~\cite{LIGOScientific:2025jau}. }
\label{tab:H0_shifts}
\end{table}

\vspace{0.5em}
\paragraph{Relating the $H_0$ posterior shift to increased freedom in mass scales.} To investigate the origin of the $H_0$ posterior shift in the redshift-evolving case, we repeat the analysis  restricting the allowed redshift evolution and the number of evolving parameters of the mass function.

We first impose tight priors on all the evolution
parameters so that the model reduces to an effectively nonevolving population.
We do so by setting $\{\sigma_{\Delta\alpha_1} = 0.2, \sigma_{\Delta\alpha_2}=0.2, \sigma_{\Delta\mu_1}=0.5, \sigma_{\Delta\sigma_1}=0.2, \sigma_{\Delta\mu_2}=1, \sigma_{\Delta\sigma_2}=1, \sigma_{\Delta m_b} = 0.5 \}$.

Second, we impose tight evolution priors on selected features while leaving the other parameters evolve. 
Specifically, we fix separately the evolution priors on: (i) mean and standard deviation of the first peak $\mu_1, \sigma_{1}$ ($\sigma_{\Delta \mu_1} = 0.2$, $\sigma_{\Delta \sigma_{1}} = 0.2$); (ii) mean and standard deviation of the second peak $\mu_2, \sigma_{2}$ ($\sigma_{\Delta \mu_2} = 1$, $\sigma_{\Delta \sigma_{2}} = 1$); (iii) power-law slopes $\alpha_1, \alpha_{2}$ ($\sigma_{\Delta \alpha_{1,2}} = 0.2$). In all these cases we also restrict the evolution for the breaking point, which is assigned a prior $\sigma_{\Delta m_b} = 0.5$. 
The resulting $H_0$ posteriors are shown in Fig.~\ref{fig:H0marg}, and inferred medians and credible levels are reported in Table~\ref{tab:H0vals}.

We observe that, as the allowed evolution is suppressed, the shift in $H_0$
disappears and the posterior converges to the result obtained with the
nonevolving mass model. 
Among the fixed parameters, fixing the low-mass peak has a negligible impact on the result, while fixing the higher-mass peak removes the shift in the $H_0$ posterior. We therefore turn out attention to discuss the impact of the increased freedom allowed by letting this peak evolve with redshift.

\vspace{0.5em}
\paragraph{Increased high--mass support favors lower $H_0$.}
Indeed, the shift toward lower values of $H_0$ can be understood as a consequence
of the additional freedom introduced by the evolving mass model for the higher-mass Gaussian peak.
In the nonevolving model the characteristic scales of the
source-frame mass distribution are effectively anchored by the nearby
events, as for sources at low distance the detector-frame to source-frame masses conversion
is only weakly affected by the
cosmological parameters.

Allowing the mass distribution to evolve with redshift relaxes this
constraint. In particular, the evolving model permits the high-mass Gaussian component of
the population to increase with redshift, effectively extending the
range of allowed source-frame masses at the high-mass end.
This additional freedom interacts with the mass-redshift degeneracy of
GW observations.  
When the population model allows larger source-frame masses, high detector-frame masses can
be interpreted as originating from heavier sources at smaller redshift; a smaller inferred redshift at fixed $d_L$ corresponds to
a smaller value of the Hubble constant. Consequently, the increased
high-mass support allowed by the evolving population model can
shift the inferred cosmological solution toward lower values of $H_0$.\footnote{Interestingly, this is in line with Ref.~\cite{Mould:2026sww} which finds a slight shift of $H_0$ toward lower values in models with more significant support for the primary mass at $m_{1, \rm src}\gtrsim 60 M_{\odot}$.}

To make this explicit, we introduce a simple proxy for the effective
location of the high-mass shoulder in the source frame of the primary-mass distribution,
\begin{equation}
X_{\rm src}(z)\equiv \mu_2(z)+2\,\sigma_2(z),
\quad
\Delta X_{\rm src}(z)\equiv X_{\rm src}(z)-X_{\rm src}(0),
\label{eq:Xsrc_proxy}
\end{equation}
where $\mu_2(z)$ and $\sigma_2(z)$ are the redshift-evolved mean and
width of the high-mass Gaussian component of the primary-mass model
(cf.\ Eq.~\eqref{eq:sigmoid_evolution}).

\begin{figure}
  \centering
 \includegraphics[width=\linewidth]{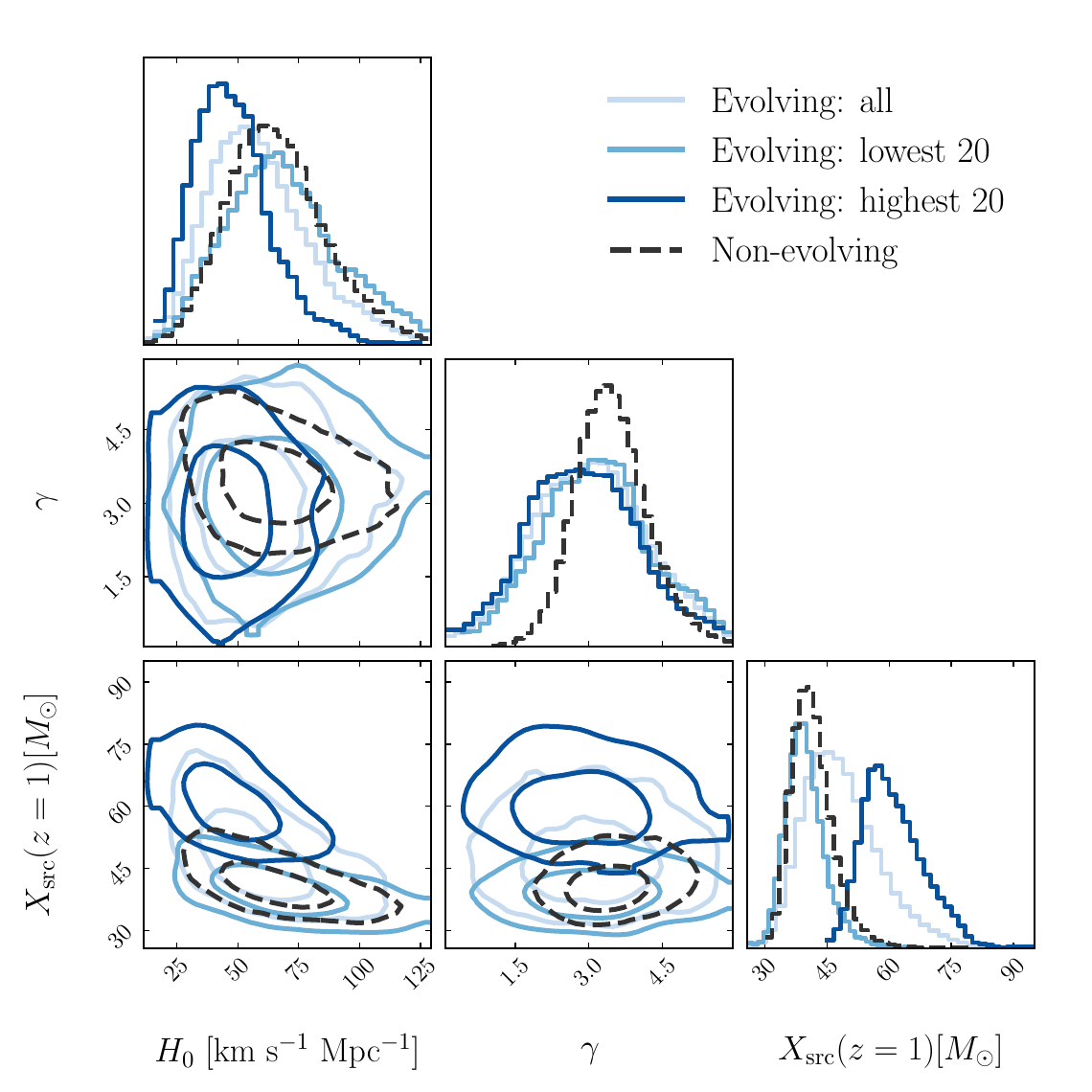}
  \caption{Posterior diagnostic of the degeneracy between cosmology and
  mass-redshift evolution in the redshift-evolving model.
  We show the joint posterior in the plane
  $\{H_0, \gamma, X_{\rm src}(z) \}$ with $X_{\rm src}(z)$ defined in
  Eq.~\eqref{eq:Xsrc_proxy}, together with slices corresponding to the
  lowest and highest 20\% quantiles of $\Delta X_{\rm src}(z)$.
  The nonevolving result is overlaid as a dashed contour.
  }
  \label{fig:H0_mass_evol_diagnostic}
\end{figure}

In Fig.~\ref{fig:H0_mass_evol_diagnostic} we show the joint posterior in
the $\{H_0, \gamma, X_{\rm src}(z=1)\}$ plane, where $\gamma$ is the low-redshift slope of the Madau-Dickinson distribution for the rate evolution with redshift.
We also slice the posterior into the lowest and highest $20\%$ quantiles
of $\Delta X_{\rm src}(z)$. Samples with larger $\Delta X_{\rm src}$
(i.e., a stronger increase of the high-mass scale with redshift)
preferentially occupy the lower-$H_0$ region of the posterior, whereas
samples with $\Delta X_{\rm src}\approx0$ yield an $H_0$ posterior close
to the nonevolving baseline. 
For reference, the contour obtained with
the nonevolving mass model is also shown as a black, dashed line. This confirms that the
low-$H_0$ shift arises from the region of parameter space in which the
mass distribution maintains or enhances high-mass support at higher
redshift. 

We also observe a systematically lower inferred value of $\gamma$ for the redshift-evolving case, namely  $\gamma =  3.0^{+1.2}_{-1.1}$ compared to $\gamma =  3.4^{+0.8}_{-0.6}$ for the nonevolving case.
A lower $\gamma$ corresponds to a weaker increase of the merger-rate density with redshift, effectively redistributing probability toward lower-redshift events. This is consistent with a lower $H_0$ values being associated to smaller inferred redshifts.
\begin{figure}
  \centering
  \includegraphics[width=\linewidth]{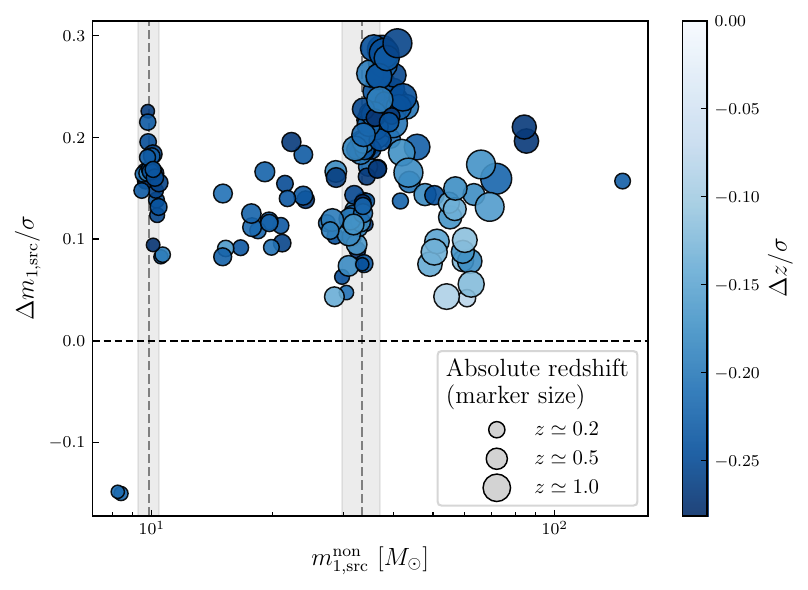}
  \caption{
  Event-level diagnostic of how allowing redshift evolution in the mass spectrum shifts the inferred
  source-frame masses and redshifts relative to the nonevolving baseline.
  Each point corresponds to one GW event.
  The horizontal axis shows the median source-frame primary mass inferred in the nonevolving model,
  $m_{1,\rm src}^{\rm non}$, while the vertical axis shows the normalized shift in the median inferred mass between the
  evolving and nonevolving models.
  The color encodes the corresponding normalized redshift shift, and marker size encodes the absolute redshift $z^{\rm non}$.
  Shaded vertical bands indicate the locations of the two main peaks of the nonevolving primary-mass spectrum.
  }
  \label{fig:event_shift_diagnostic}
\end{figure}

\vspace{0.5em}
\paragraph{Event--level diagnostics.}
A complementary view of the mass-cosmology degeneracy can be obtained at the level
of individual events.
Starting from the hyperposterior samples of the hierarchical inference, we reconstruct
\emph{population-informed} posteriors for each event by reweighting the event-level posterior samples
according to the inferred population model (a standard procedure in hierarchical population analyses).
For each event $n$, this yields population-reweighted medians for the evolving model,
$(z_n^{\rm evol}, m_{1,n}^{\rm evol})$, and for the nonevolving model,
$(z_n^{\rm non}, m_{1,n}^{\rm non})$, which can be compared directly.

In Fig.~\ref{fig:event_shift_diagnostic} we visualize these shifts in the plane
$\{m_{1,\rm src}^{\rm non}, \Delta m_{1,\rm src}\}$, where
\begin{equation}
\frac{\Delta m_{1,\rm src}}{\sigma_{m_{1, \rm{src}}}}=
\frac{m_{1,\rm src}^{\rm evol}-m_{1,\rm src}^{\rm non}}
{\sqrt{\left(\sigma_{m_1}^{\rm evol}\right)^2+\left(\sigma_{m_1}^{\rm non}\right)^2}}
\end{equation}
is the shift in the median source-frame primary mass, normalized by the combined posterior width of the two.
The point color encodes the corresponding standardized redshift shift
\begin{equation}
\frac{\Delta z}{\sigma_z} =
\frac{z^{\rm evol}-z^{\rm non}}
{\sqrt{\left(\sigma_{z}^{\rm evol}\right)^2+\left(\sigma_{z}^{\rm non}\right)^2}},
\end{equation}
so that negative values (darker shades) indicate a reduction in the inferred redshift when evolution is allowed.
Marker size encodes the absolute redshift inferred under the nonevolving model, $z_n^{\rm non}$.
Vertical shaded bands indicate the locations of the two main peaks of the nonevolving primary-mass spectrum.

A clear qualitative trend is that allowing redshift evolution induces a coherent shift across the catalog:
for most events, the inferred source-frame mass increases ($\Delta m_{1,\rm src}>0$) while the inferred redshift
decreases ($\Delta z<0$).\footnote{Two low-mass events (\texttt{GW190924\_021846} and \texttt{GW230627\_015337}) present a negative mass shift. This arises because population reweighting slightly shifts their detector-frame posterior toward lower $m_{1,\rm det}$ in the evolving fit, and this change is large enough to outweigh the modest decrease in $(1+z)$, yielding $\Delta z<0$ and $\Delta m_{1,\rm src}<0$ simultaneously.}
This is the event-level signature expected from the spectral-siren mapping
$m_{\rm src}=m_{\rm det}/(1+z)$: reducing the inferred redshift at fixed detector-frame mass necessarily raises the inferred
source-frame mass.
In the evolving model, allowing evolution enlarges the space of admissible high-mass support at moderate-to-high redshift, so that the same detector-frame masses can be accommodated with slightly smaller inferred redshifts.
The effect is most visible for events associated with the higher-mass Gaussian component near $m_{1,\rm src}\sim 35\,M_\odot$:
because this feature lies at larger masses (and is therefore preferentially populated by higher-$z$ detections), additional freedom in its location/width and relative weight has a larger impact on the inferred redshifts and hence on the cosmology-population degeneracy.
This catalog-wide tendency toward smaller inferred redshifts when evolution is allowed is the same direction preferred by lower $H_0$, providing an event-level perspective on the shift in the $H_0$ posterior observed in the evolving model.

\begin{figure*}[t]
\centering
\includegraphics[width=0.5\linewidth]{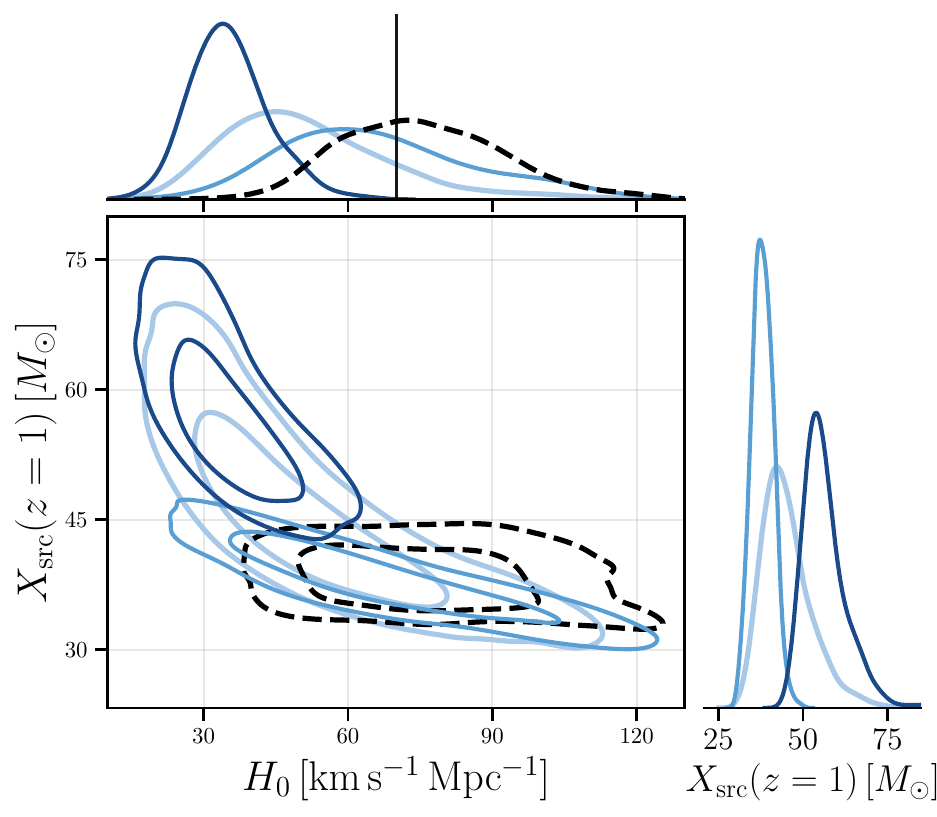}
\includegraphics[width=0.46\linewidth]{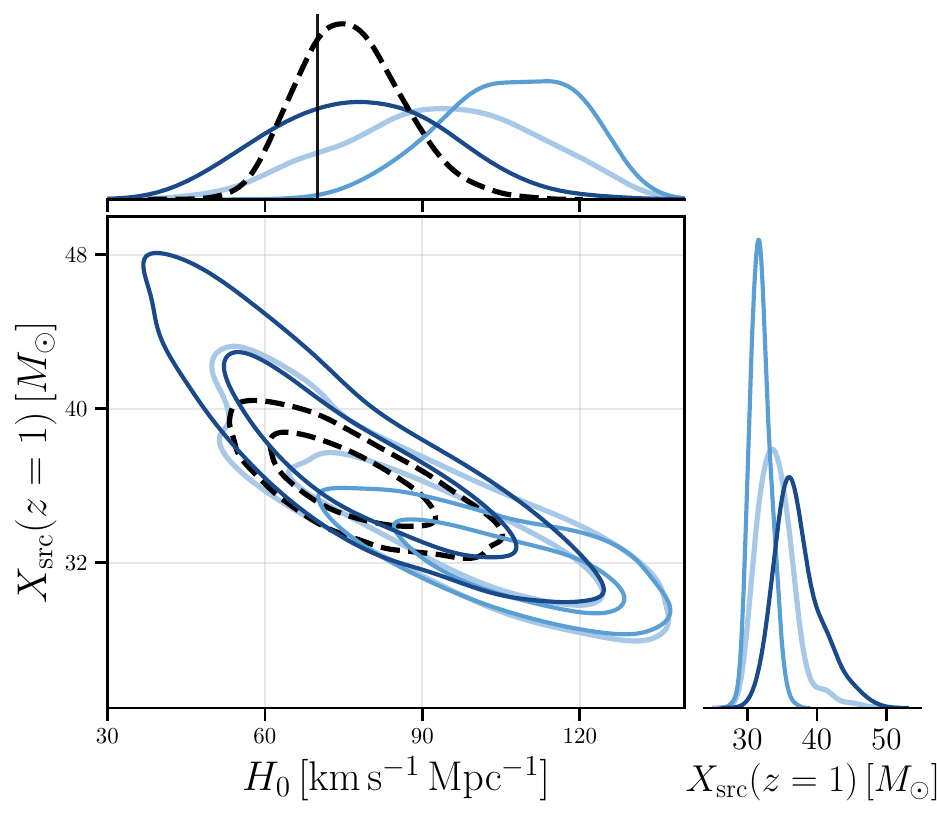}
\caption{
    Analog to Fig.~\ref{fig:H0_mass_evol_diagnostic}, for a simulated scenario where the underlying model is nonredshift-evolving. 
    We show the joint posterior in the plane
  $\{H_0, X_{\rm src}(z=1) \}$, with slices corresponding to the
  lowest and highest 20\% quantiles of $\Delta X_{\rm src}(z)$.
  Dashed contours corresponds to an analysis that assumes a nonevolving model.
Left: 150 detections at O4-like sensitivity. Right: 1000 detections at O5-like sensitivity.
}
\label{fig:H0_mass_evol_diagnostic_sim}
\end{figure*}

\subsection{Comparison with simulations}\label{sec:sim}

Finally, we investigate whether the observed trend in the $H_0$ posterior, together with the absence of evidence for evolution, can be coherently reproduced with simulations. 

To this end, we generate simulated catalogs of BBH mergers assuming the same mass and redshift distributions used in the nonevolving population model described in Sec.~\ref{sec:population_model}, namely the default \textsc{Broken Power Law + 2 Peaks} model of the LVK GWTC-4.0 population analysis~\cite{LIGOScientific:2025pvj}.  
The corresponding fiducial hyperparameter values adopted for the simulations are the median values of the LVK posterior samples available at Ref.~\cite{ligo_scientific_collaboration_2025_16911563}. 
For simplicity, we neglect spin degrees of freedom in the simulated sources.
We use a modified version of the \texttt{GWMockCat} \texttt{python} package~\cite{Farah:2023vsc} to generate mock data and injections sets. 

The mock data are then analyzed with a redshift-evolving model. 
Figure~\ref{fig:H0_mass_evol_diagnostic_sim} shows the joint posterior in
the $\{H_0, X_{\rm src}(z_\ast=1)\}$ plane for an LVK O4-like observing scenario, with $150$ detected events (left), and an LVK O5-like observing scenario, with $1000$ detected events (right), all with $\text{SNR}>8$.

At O4 sensitivity, the observed trend is fully consistent with that found for GWTC-4.0 (Fig.~\ref{fig:H0_mass_evol_diagnostic}): samples in the lowest quantile of $\Delta X_{\rm src}$ are associated with values of $H_0$ consistent with the nonevolving (correct) baseline, while those in the upper tail tend to shift $H_0$ toward lower values.

Interestingly, if this interpretation is correct, the trend at O5 sensitivity becomes partially reversed. In the right panel of Fig.~\ref{fig:H0_mass_evol_diagnostic_sim}, the overall $H_0$ posterior shifts toward higher values, driven by increased support from the low-$20\,\%$ quantile tail of $\Delta X_{\rm src}$.
This behavior likely reflects the increased statistical constraining power of the data at higher redshift in the O5 scenario. In particular, the position of the $\sim35\,M_\odot$ feature becomes increasingly constrained directly by the data, rather than by the prior. Since the underlying simulated population is nonevolving, the low-redshift events anchor the source-frame position of the peak, reducing the freedom of the evolving model to reinterpret high detector-frame mass events as intrinsically more massive systems at lower redshift.
Instead, high detector-frame masses are more naturally explained as standard source-frame masses observed at larger redshift through the relation $m_{\rm det}=(1+z)\,m_{\rm src}$. This increases the posterior support for configurations with larger inferred distances and therefore larger values of $H_0$.

Overall, these results suggest that, at current sensitivity, the observed trend can be explained by a nonredshift-evolving underlying mass spectrum (at least within the redshift reach of current observations) combined with the larger freedom allowed by the redshift-evolving assumption. Should this remain true at higher redshifts, the trend in the $H_0$ shift may change as detector sensitivity increases.

We stress that these conclusions depend on the assumed underlying generative model; more observations will be required to discriminate between scenarios, and targeted diagnostics will become increasingly important as the catalog size and redshift reach grow.
In particular, the purpose of this injection study is not to demonstrate that the observed catalog is generated by a stationary mass distribution. Since the injected population is nonevolving by construction, the comparison only shows that a mild displacement of the $H_0$ posterior similar to the one observed in real data can arise from fitting an overflexible evolving model to a stationary population at current sensitivity. Distinguishing this scenario from that of a mildly evolving underlying population requires dedicated injections with controlled redshift evolution, which we leave to future work. We therefore interpret the mock-catalog result only as evidence that the real-data shift is compatible with overflexibility, not as evidence that the real data are stationary.

\section{Conclusions}\label{sec:conclusions}

Spectral-siren measurements of the Hubble constant rely on identifying features in the source-frame BBH mass spectrum and using them as statistical redshift anchors. This makes the inference potentially sensitive to assumptions about the population model, and in particular to possible redshift evolution of the mass spectrum. In this work, we revisited GWTC-4.0 spectral-siren cosmology by explicitly allowing the main parameters of a \textsc{Broken Power Law + 2 Peaks} BBH mass model to evolve with redshift.

We first analyzed the catalog at fixed cosmology and found no conclusive evidence for redshift evolution in the primary-mass spectrum. Posterior predictive checks show that both the nonevolving and redshift-evolving models provide statistically adequate descriptions of the observed catalog. 
A KL-divergence-based comparison between the inferred $p(m_1\mid z)$ distributions at $z=0.2$ and $z=1$ similarly indicates consistency with no overall evolution at $\sim 93\%$. 
At the level of localized mass scales, the $95\%$ credible interval of the ratio $p(m_1\mid z=1)/p(m_1\mid z=0.2)$ remains consistent with no evolution across the full mass range, although some localized deviations appear at the $68\%$ level around the Gaussian peak features. These deviations are not statistically significant and are consistent with the additional prior volume introduced by the evolving model.

When allowing cosmology to vary, we find that the redshift-evolving model induces only a small shift of the inferred $H_0$ posterior toward lower values relative to the nonevolving baseline. However, this shift is far from being statistically significant, corresponding to only $\simeq 0.3\,\sigma$ when expressed in units of the combined uncertainty of the two measurements. We therefore conclude that, within the flexibility considered here, redshift evolution of the mass spectrum does not represent a dominant systematic for GWTC-4.0 spectral-siren measurements of $H_0$.

We compared this shift with recent GWTC-4.0 spectral-siren studies that investigate alternative redshift-independent descriptions of the mass spectrum, including additional peaks, sharper spectral features, and spline-based reconstructions. The shifts induced by those modeling choices are generally comparable to, and often larger than, the shift associated here with allowing redshift evolution. 
This suggests that, at current sensitivity, population-model flexibility associated with redshift-independent spectral features likely represents a more important source of uncertainty for spectral-siren cosmology than smooth redshift evolution of the mass spectrum.

To understand the origin of the residual $H_0$ shift, we introduced posterior and event-level diagnostics. These show that the low-$H_0$ tail is associated with posterior samples in which the high-mass Gaussian component gains additional support at higher redshift. In such samples, high detector-frame masses are preferentially interpreted as intrinsically more massive systems at lower inferred redshift through the relation $m_{\rm det}=(1+z)m_{\rm src}$, which in turn favors lower values of $H_0$. Event-level reweighting confirms this picture: when evolution is allowed, many events shift coherently toward larger source-frame masses and smaller inferred redshifts, consistent with the direction of the observed cosmological shift.

Finally, we used simulated catalogs to test whether the observed behavior can arise even when the underlying population is nonevolving. In O4-like simulations, analyzing a nonevolving population with a redshift-evolving model reproduces the qualitative trend observed in GWTC-4.0, suggesting that the current shift may be caused by fitting an overly flexible model to a catalog that does not yet require such freedom. At higher sensitivity and larger redshift reach, however, the sign and magnitude of the induced shift may change, as illustrated by simulations in an O5-like observing scenario, as the increased statistical power of higher-redshift events progressively constrains the location of the spectral features directly from the data. 

Overall, our results indicate that GWTC-4.0 spectral-siren constraints on $H_0$ are robust to the smooth redshift evolution of the main spectral features of the primary BBH mass spectrum. We find no statistically significant evidence for such evolution, and the small induced shift in $H_0$ is comparable to, if not subdominant to, other population-modeling uncertainties already explored in the literature. Future catalogs with larger event numbers and greater redshift reach will be essential to determine whether the apparent localized trends in the mass spectrum develop into significant evolution, and to disentangle genuine astrophysical evolution from modeling flexibility in spectral-siren cosmology.

\acknowledgements

We thank Matteo Tagliazucchi for discussions, Michele Dell'Aquila for valuable mathematical insights, and Utkarsh Mali for internal LIGO-Virgo review and useful comments on the draft.
This work is supported by the French government under the France 2030 investment plan, as part of the Initiative d'Excellence d'Aix-Marseille Universit\'e---A*MIDEX AMX-22-CEI-02.

{\small This research has made use of data or software obtained from the Gravitational Wave Open Science Center (gwosc.org), a service of the LIGO Scientific Collaboration, the Virgo Collaboration, and KAGRA. This material is based upon work supported by NSF's LIGO Laboratory which is a major facility fully funded by the National Science Foundation, as well as the Science and Technology Facilities Council (STFC) of the United Kingdom, the Max-Planck-Society (MPS), and the State of Niedersachsen/Germany for support of the construction of Advanced LIGO and construction and operation of the GEO600 detector. Additional support for Advanced LIGO was provided by the Australian Research Council. Virgo is funded, through the European Gravitational Observatory (EGO), by the French Centre National de Recherche Scientifique (CNRS), the Italian Istituto Nazionale di Fisica Nucleare (INFN) and the Dutch Nikhef, with contributions by institutions from Belgium, Germany, Greece, Hungary, Ireland, Japan, Monaco, Poland, Portugal, Spain. KAGRA is supported by Ministry of Education, Culture, Sports, Science and Technology (MEXT), Japan Society for the Promotion of Science (JSPS) in Japan; National Research Foundation (NRF) and Ministry of Science and ICT (MSIT) in Korea; Academia Sinica (AS) and National Science and Technology Council (NSTC) in Taiwan.
}

This document has LIGO document No. LIGO-P2600238.

\section*{Data availability}
The code used for this work is available at~Ref.~\cite{pymcpop}. 
Version v1.0.0, used for this paper, is archived on Zenodo at Ref.~\cite{michele_mancarella_2026_21837861}.
Data products for events in the GWTC-4.0 catalog are available at~\cite{ligo_scientific_collaboration_and_virgo_2025_17014085, ligo_scientific_collaboration_and_virgo_2023_8177023, ligo_scientific_collaboration_and_virgo_2022_6513631}.
Sensitivity estimates are available at~\cite{ligo_scientific_collaboration_2025_16740128}.

\bibliography{correlations}

\appendix

\appendix
\section{Complete population model specification}
\label{app:population_model}

This Appendix provides the full mathematical definition of the population
model summarized in Sec.~\ref{sec:population_model}.  In particular we give
explicit expressions for the conditional mass distributions, tapering
functions, redshift distribution.

\paragraph{Primary-mass distribution}

At fixed redshift the primary-mass distribution is given by a mixture
between a broken power-law component and two Gaussian components as in Eq.~\ref{eq:mixture_pm1}, namely:
\begin{equation}
\begin{split}
p(m_1 \mid z, \Lambda_{m_1})
\propto
& \, T(m_1) \,\Big[
\lambda_0(z)\,p_{\mathrm{BPL}}(m_1 \mid z)
+ \\
& \lambda_1(z)  p_{\mathrm{G},1}(m_1 \mid z)
+ \lambda_2(z)\,p_{\mathrm{G},2}(m_1 \mid z)
\Big]\,
,
\end{split}
\label{eq:mixture_pm1_app}
\end{equation}

The broken power-law component is defined as
\begin{equation}
p_{\rm BPL}(m_1\mid z)\propto
\begin{cases}
\left(\dfrac{m_1}{m_b}\right)^{-\alpha_1(z)}, & m_1 < m_b, \\
\left(\dfrac{m_1}{m_b}\right)^{-\alpha_2(z)}, & m_1 \ge m_b ,
\end{cases}
\end{equation}
where $m_b$ denotes the break mass.

The Gaussian components are given by
\begin{equation}
p_{{\rm G},i}(m_1\mid z)
\propto
\exp\!\left[
-\frac{(m_1-\mu_i(z))^2}{2\sigma_i^2(z)}
\right],
\qquad i=1,2,
\end{equation}
each truncated and renormalized over the interval
$[m_{1,\min},m_{1,\max}]$.

\paragraph{Secondary-mass distribution}

The secondary mass $m_2$ is modeled through a power-law distribution
conditioned on the primary mass,
\begin{equation}
p(m_2\mid m_1,\Lambda_{m_2})
\propto
m_2^{\beta}\,T(m_2)\,\mathcal{H}(m_1-m_2),
\end{equation}
where $\mathcal{H}$ is the Heaviside step function enforcing the
ordering $m_2\le m_1$. For each value of $m_1$, the distribution is
normalized over the interval $m_2\in[m_{2,\min},m_1]$.

\paragraph{Low-mass tapering}

To avoid discontinuities associated with hard cutoffs, we apply a smooth tapering function to both component masses.
For a mass variable $m$ with minimum value $m_{\min}$ and taper width
$\delta m$, we define
\begin{equation}
t(m)=\frac{m-m_{\min}}{\delta m},
\end{equation}
and the taper function
\begin{equation}
T(m)=
\begin{cases}
0, & t\le 0, \\
3t^2-2t^3, & 0<t<1, \\
1, & t\ge 1.
\end{cases}
\end{equation}
This cubic smoothstep function is $C^1$ continuous, monotonic,
and has vanishing derivative at both endpoints of the transition.

Note that this functional form differs from the standard LVK choice, see Ref.~\cite{LIGOScientific:2025pvj}, and is chosen to improve gradient calculation in Hamiltonian Monte Carlo.

The primary and secondary masses use the pairs
$(m_{1,\min},\delta m_1)$ and $(m_{2,\min},\delta m_2)$,
respectively.

\paragraph{Redshift distribution}

The redshift distribution is modeled as
\begin{equation}
p(z\mid\Lambda_z)
\propto
\frac{dV_c}{dz}\,
\frac{\psi(z\mid\Lambda_z)}{1+z},
\end{equation}
where $dV_c/dz$ is the differential comoving volume element
in a flat $\Lambda$CDM cosmology and the factor $(1+z)^{-1}$
accounts for cosmological time dilation.

We adopt a Madau--Dickinson merger-rate density
\cite{Madau:2014bja},
\begin{equation}
\psi(z\mid\gamma,\kappa,z_p)
\propto
\frac{(1+z)^\gamma}
{1+\left(\dfrac{1+z}{1+z_p}\right)^{\gamma+\kappa}},
\end{equation}
which reproduces the rise of the cosmic star-formation rate at
low redshift and its decline above the peak redshift $z_p$.

\section{Priors and prior predictive checks}
\label{app:priors}

This Appendix provides the detailed prior implementation corresponding to the
model described in Sec.~\ref{sec:methods}, including (i) the explicit sampling
parametrizations used for bounded parameters and redshift-evolution triplets,
(ii) the prior ranges for each parameter, and (iii) prior predictive checks for
the implied primary-mass density $p(m_1\mid z)$.

\subsection{Implementation details}

\paragraph{Bounded-sigmoid priors (raw-space sampling).}
For any parameter $\phi$ with prescribed bounds $[\phi_{\min},\phi_{\max}]$ we sample an
unconstrained latent variable $\phi_{\rm raw}$ and map it to the bounded domain:
\begin{align}
\phi_{\rm raw} &\sim \mathcal{N}(0,\sigma_{\rm raw}),\\
\phi &= \phi_{\min} + (\phi_{\max}-\phi_{\min})\,\mathrm{sigmoid}(x_{\rm raw}).
\end{align}
The scale $\sigma_{\rm raw}$ controls concentration around the midpoint of the interval. We use $\sigma_{\rm raw}=1$ for obtaining a mildly informative prior centered near $(\phi_{\min}+\phi_{\max})/2$,
and $\sigma_{\rm raw}=1.5$ for an approximately flat prior over the interior with
softened edges.
We use this construction for most bounded parameters in the analysis
(including cosmology and several population hyperparameters).

\paragraph{Log-bounded variant for scale parameters.}
When a parameter spans orders of magnitude and we wish the prior to be symmetric around
a reference value in logarithmic space, we apply the same
bounded-sigmoid construction to $\ln \phi$ instead of $\phi$:
\begin{align}
\ln x_{\rm raw} &\sim \mathcal{N}(0,\sigma_{\rm raw}),\\
\ln \phi &= \ln \phi_{\min} + (\ln \phi_{\max}-\ln \phi_{\min})\,\mathrm{sigmoid}\big(\ln x_{\rm raw}\big).
\end{align}

\paragraph{Redshift-evolution triplets.}
For each evolving hyperparameter $\phi\in\{\alpha_1,\alpha_2, m_b, \mu_1,\sigma_1,\mu_2,\sigma_2\}$ we sample
\begin{align}
\Delta\phi &\sim \mathcal{N}(0,\sigma_{\Delta\phi}), \\
\phi_\infty &= \phi_0 + \Delta\phi,\\
z_\phi &\sim \mathrm{Uniform}(z_{\min}, z_{\max}),\\
\log \Delta z_\phi &\sim \mathcal{N}(\log \Delta z_{\rm mid}, \sigma_{\log \Delta z}).
\end{align}
For parameters constrained to remain above a physical floor (e.g.\
$\sigma_i>0$), we impose the constraint at the level of $\Delta\phi$ via a
one-sided truncation such that $\phi_\infty>\epsilon_{\rm pos}$.

\paragraph{Auxiliary parametrization for low-mass bounds}
\label{app:bounds_nonevol}

The low-mass bound of the primary mass is sampled via an auxiliary variable $u\equiv \mathrm{sigmoid}(u_{\rm raw}) \in(0,1)$:
\begin{equation}
\begin{split}
& m_{1,\min}=3 + (10-3)\,\left[\mathrm{sigmoid}(u_{\rm raw})\right]^{3/2}, \\
& u_{\rm raw}\sim \mathcal{N}(0,1),
\end{split}
\end{equation}
so that $m_{1,\min}\in[3,10]\,M_\odot$ while preserving a smooth unconstrained sampling geometry.
The secondary low-mass bound is constructed conditionally using $v\equiv \mathrm{sigmoid}(v_{\rm raw}) \in(0,1)$:
\begin{equation}
\begin{split}
& m_{2,\min}=3 + \mathrm{sigmoid}(v_{\rm raw})\,(m_{1,\min}-3), \\
& v_{\rm raw}\sim \mathcal{N}(0,1),
\end{split}
\end{equation}

which enforces $m_{2,\min}\le m_{1,\min}$ by construction.

Table~\ref{tab:priors} summarizes the priors used in the analysis.

\begin{table*}
\centering
\footnotesize
\begin{tabular}{llll}
\toprule
Parameter & Sampling prior (in raw space) & Transform & Equivalent prior bounds \\
\midrule
\multicolumn{4}{c}{\emph{Cosmology and propagation}}
\\
\\
$H_0$ &
$H_{0,\rm raw}\sim \mathcal{N}(0,\sigma_{\rm raw}=1.5)$ &
$H_0 = 10 + (130-10)\,\mathrm{sigmoid}(H_{0,\rm raw})$ &
$H_0\in[10,130]$ $\mathrm{km\,s^{-1}\,Mpc^{-1}}$\\

$\Omega_m$ &
$\Omega_{m,\rm raw}\sim \mathcal{N}(0,\sigma_{\rm raw}=1)$ &
$\Omega_m = 0.05 + (0.55-0.05)\,\mathrm{sigmoid}(\Omega_{m,\rm raw})$ &
$\Omega_m\in[0.05,0.55]$ \\

\midrule

\multicolumn{4}{c}{\emph{Merger-rate density (Madau--Dickinson type)}}
\\
\\
$\gamma$ &
$\gamma_{\rm raw}\sim \mathcal{N}(0,\sigma_{\rm raw}=1.5)$ &
$\gamma = 6\,\mathrm{sigmoid}(\gamma_{\rm raw})$ &
$\gamma\in[0,6]$ \\

$\kappa$ &
$\kappa_{\rm raw}\sim \mathcal{N}(0,\sigma_{\rm raw}=1)$ &
$\kappa = 6\,\mathrm{sigmoid}(\kappa_{\rm raw})$ &
$\kappa\in[0,6]$ \\

$z_p$ &
$z_{p,\rm raw}\sim \mathcal{N}(0,\sigma_{\rm raw})$ &
$z_p = 4\,\mathrm{sigmoid}(z_{p,\rm raw})$ &
$z_p\in[0,4]$ \\

\midrule

\multicolumn{4}{c}{\emph{Primary-mass model at $z\simeq 0$ \textsc{BPL + 2P}}}
\\
\\
$(\alpha_{1,0},\alpha_{2,0})$ &
\begin{tabular}[t]{@{}l@{}}
$\alpha_{\rm bar}\sim\mathcal{N}(a_{\rm mid},a_\sigma),\;
\alpha_{\rm diff}\sim\mathcal{N}(0,\sqrt{2}\,a_\sigma)$\\
$a_{\rm mid}=\frac{1.5+7}{2},\;
a_\sigma=\frac{7-1.5}{2\times 1.95996}$
\end{tabular} &
$\alpha_{1,0}=\alpha_{\rm bar}-\frac{1}{2}\alpha_{\rm diff},\;
\alpha_{2,0}=\alpha_{\rm bar}+\frac{1}{2}\alpha_{\rm diff}$ &
$\alpha_{1,0},\alpha_{2,0}\in[1.5,7]$ \\

$m_{b,0}$ &
$m_{b,\rm raw}\sim\mathcal{N}(0,\sigma_{\rm raw})$ &
$m_{b,0} = 20 + (50-20)\,\mathrm{sigmoid}(m_{b,\rm raw})$ &
$m_{b,0}\in[20,50]$ \\

$\mu_{1,0}$ &
$\mu_{1,\rm raw}\sim\mathcal{N}(0,\sigma_{\rm raw})$ &
$\mu_{1,0} = 5 + (15-5)\,\mathrm{sigmoid}(\mu_{1,\rm raw})$ &
$\mu_{1,0}\in[5,15]$ \\

$\sigma_{1,0}$ &
$\sigma_{1,\rm tail}\sim\mathrm{LogNormal}(\mu_{\ln},\sigma_{\ln})$ &
$\sigma_{1,0}=0.1+\sigma_{1,\rm tail}$ &
$\sigma_{1,0}\in[0.1,5]$ \\

$\mu_{2,0}$ &
$\mu_{2,\rm raw}\sim\mathcal{N}(0,\sigma_{\rm raw})$ &
$\mu_{2,0} = 25 + (55-25)\,\mathrm{sigmoid}(\mu_{2,\rm raw})$ &
$\mu_{2,0}\in[25,55]$ \\

$\sigma_{2,0}$ &
$\sigma_{2,\rm tail}\sim\mathrm{LogNormal}(\mu_{\ln},\sigma_{\ln})$ &
$\sigma_{2,0}=0.5+\sigma_{2,\rm tail}$ &
$\sigma_{2,0}\in[0.5,10]$ \\

$m_{1,\min}$ &
$u_{\rm raw}\sim\mathcal{N}(0,1)$ &
$u=\mathrm{sigmoid}(u_{\rm raw}),\;
m_{1,\min}=3+(10-3)\,u^{1.5}$ &
$m_{1,\min}\in[3,10]$ \\

$m_{1,\max}$ &
$\Delta m_{\rm high}\sim \mathrm{LogNormal}(\mu,\sigma)$ &
$m_{1,\max}=100+\Delta m_{\rm high}$ &
$m_{1,\max}\in[100,250]$ \\

$\delta m_1$ &
$\delta m_{1,\rm tail}\sim\mathrm{LogNormal}(\mu_{\ln},\sigma_{\ln})$ &
$\delta m_1=0.1+\delta m_{1,\rm tail}$ &
$\delta m_1\in[0.1,10]$ \\

$\boldsymbol{\lambda}_0$ &
$\boldsymbol{\lambda}_0\sim\mathrm{Dirichlet}(1,1,1)$ &
(simplex-valued) &
$\boldsymbol{\lambda}_0\in\Delta^2$ \\

\midrule

\multicolumn{4}{c}{\emph{Secondary-mass model}}
\\
\\

$\beta$ &
$\beta\sim\mathcal{N}(\mu,\sigma)$ &
$\mu=\frac{-2+7}{2},\;
\sigma=\frac{7-(-2)}{2\times 1.95996}$ &
$\beta\in[-2,7]$ \\

$m_{2,\min}$ &
$v_{\rm raw}\sim\mathcal{N}(0,1)$ &
$v=\mathrm{sigmoid}(v_{\rm raw}),\;
m_{2,\min}=3+v(m_{1,\min}-3)$ &
$m_{2,\min}\in[3,m_{1,\min}]$ \\

$\delta m_2$ &
$\delta m_{2,\rm tail}\sim\mathrm{LogNormal}(\mu_{\ln},\sigma_{\ln})$ &
$\delta m_2=0.1+\delta m_{2,\rm tail}$ &
$\delta m_2\in[0.1,10]$ \\

\midrule

\multicolumn{4}{c}{\emph{Redshift evolution}}
\\
\\
$\Delta\alpha_1$ &
$\mathcal{N}(0,\sigma_{\Delta\alpha_1}=2.0)$ &
$\alpha_{1,\infty}=\alpha_{1,0}+\Delta\alpha_1$ &
$\alpha_{1,\infty}\in[-2.42,10.92]$ \\

$\Delta\alpha_2$ &
$\mathcal{N}(0,\sigma_{\Delta\alpha_2}=1.5)$ &
$\alpha_{2,\infty}=\alpha_{2,0}+\Delta\alpha_2$ &
$\alpha_{2,\infty}\in[-1.44,9.94]$  \\

$\Delta m_b$ &
$\mathcal{N}(0,\sigma_{\Delta m_b}=10.0)$, truncated so $m_{b,\infty}>5\,M_\odot$ &
$m_{b,\infty}=m_{b,0}+\Delta m_b$ &
$m_{b,\infty}\in[5,69.60]\,M_\odot$ \\

$\Delta\mu_1$ &
$\mathcal{N}(0,\sigma_{\Delta\mu_1}=5.0)$, truncated so $\mu_{1,\infty}>5\,M_\odot$ &
$\mu_{1,\infty}=\mu_{1,0}+\Delta\mu_1$ &
$\mu_{1,\infty}\in[5,24.80]\,M_\odot$  \\

$\Delta\sigma_1$ &
$\mathcal{N}(0,\sigma_{\Delta\sigma_1}=3.0)$, truncated so $\sigma_{1,\infty}>0.1$ &
$\sigma_{1,\infty}=\sigma_{1,0}+\Delta\sigma_1$ &
$\sigma_{1,\infty}\in[0.1,10.88]\,M_\odot$  \\

$\Delta\mu_2$ &
$\mathcal{N}(0,\sigma_{\Delta\mu_2}=10.0)$, truncated so $\mu_{2,\infty}>5\,M_\odot$ &
$\mu_{2,\infty}=\mu_{2,0}+\Delta\mu_2$ &
$\mu_{2,\infty}\in[5,74.60]\,M_\odot$  \\

$\Delta\sigma_2$ &
$\mathcal{N}(0,\sigma_{\Delta\sigma_2}=10.0)$, truncated so $\sigma_{2,\infty}>0.1$ &
$\sigma_{2,\infty}=\sigma_{2,0}+\Delta\sigma_2$ &
$\sigma_{2,\infty}\in[0.1,29.60]\,M_\odot$ \\

$z_\phi$ & $\mathrm{Uniform}(z_{\min},z_{\max})$ & used in Eq.~\ref{eq:sigmoid_evolution} & $z_\phi\in[0.05,1.5]$ \\
$\Delta z_\phi$ & $\log\Delta z_\phi\sim\mathcal{N}(\log \Delta z_{\rm mid},0.5)$ & $\Delta z_\phi=\exp(\log\Delta z_\phi)$ & $\Delta z_\phi\in[0.05,2.0]$ \\

$\boldsymbol{\lambda}_\infty$ &
$\boldsymbol{\lambda}_\infty\sim\mathrm{Dirichlet}(1,1,1)$ &
(simplex-valued) &
$\boldsymbol{\lambda}_\infty\in\Delta^2$ \\

$z_\lambda$ & $\mathrm{Uniform}(0.05,1.5)$ & shared transition for weights & $z_\lambda\in[0.05,1.5]$ \\
$\Delta z_\lambda$ & $\log \Delta z_\lambda \sim \mathrm{Uniform}(\log 0.05,\log 2.0)$ & shared transition for weights & $\Delta z_\lambda\in[0.05,2.0]$ \\

\bottomrule
\end{tabular}

\caption{Priors used in this work, including the explicit numerical values adopted in the analysis. Reported bounds correspond either to hard prior support (for transformed bounded parameters) or to approximate 95\% prior intervals (for Gaussian/log-normal hyperparameters).}
\label{tab:priors}
\end{table*}

\begin{figure*}
\centering
\includegraphics[width=0.9\textwidth]{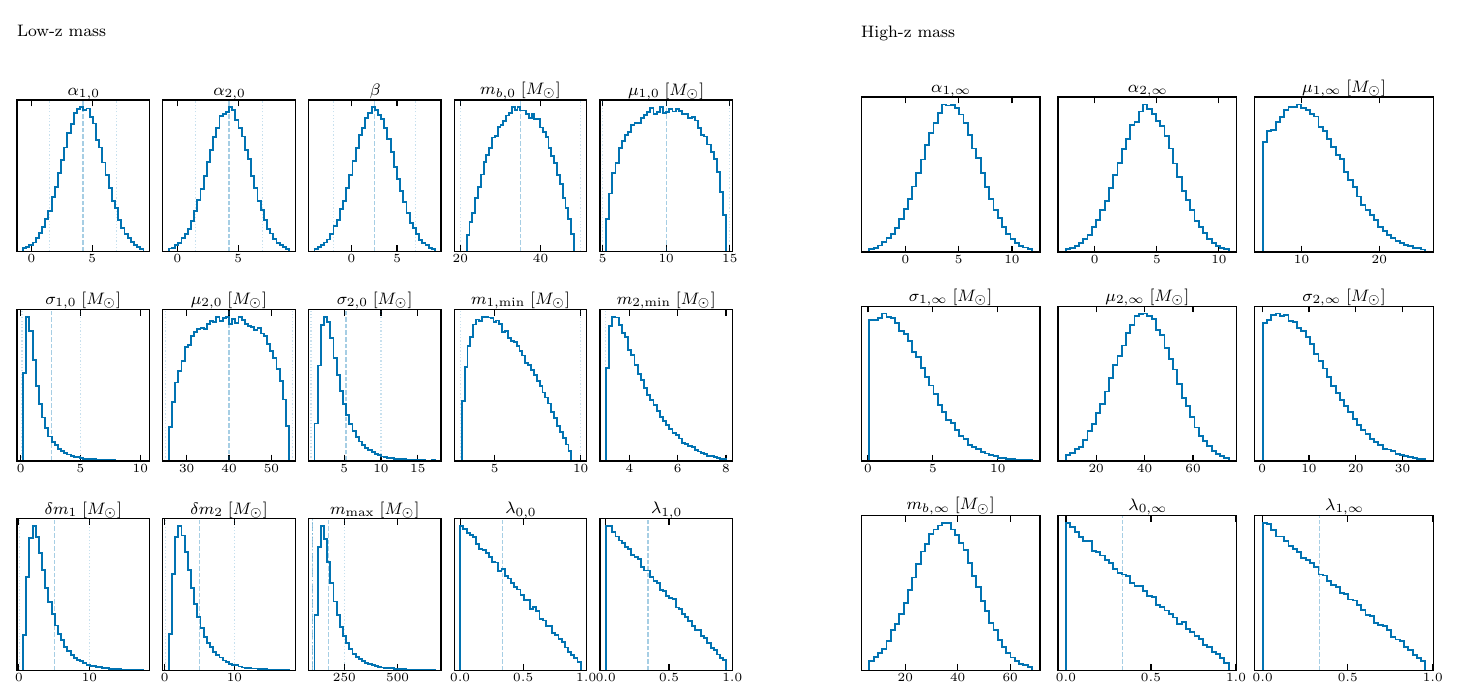}
\caption{One-dimensional prior marginals for mass model hyperparameters.}
\label{fig:prior_marginals}
\end{figure*}

\begin{figure}
\centering
\includegraphics[width=0.45\textwidth]{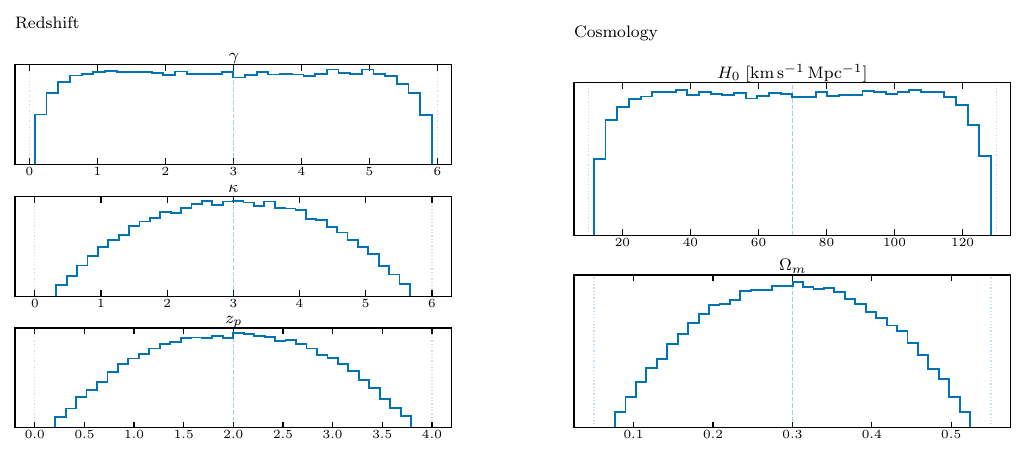}
\caption{One-dimensional prior marginals for rate and cosmology parameters.}
\label{fig:prior_marginals_1}
\end{figure}

Figures~\ref{fig:prior_marginals}-\ref{fig:prior_marginals_1} show the one-dimensional prior marginals for
all hyperparameters used in the analysis.

\begin{figure}
\centering
\includegraphics[width=0.45
\textwidth]{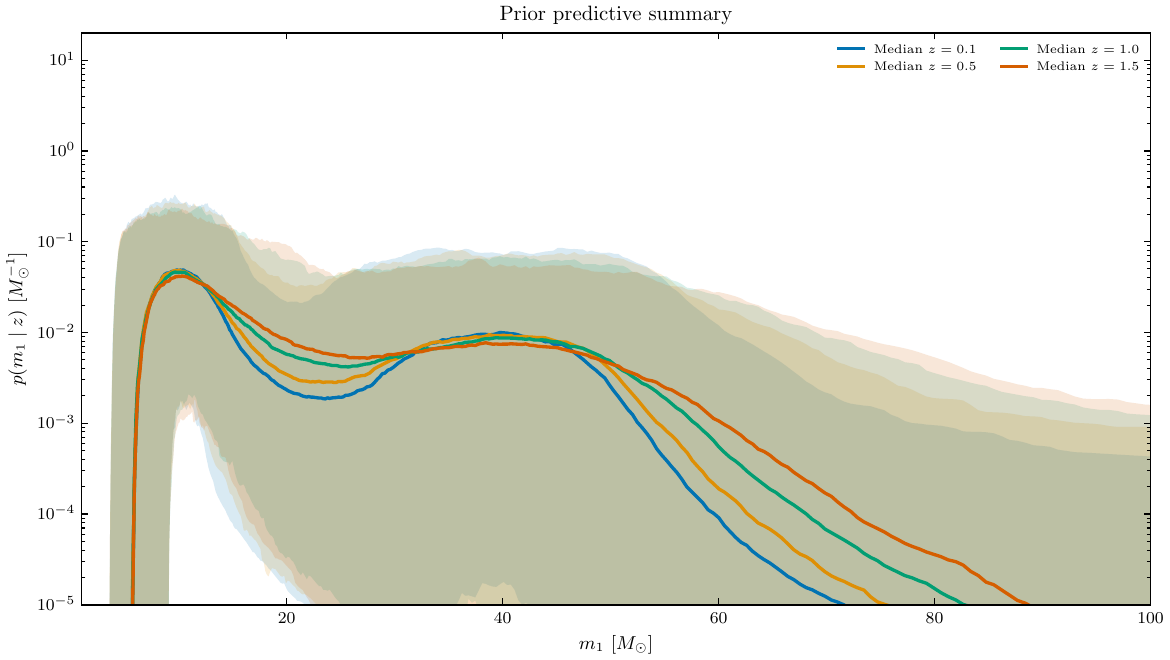}
\caption{Prior predictive distribution of the primary-mass density $p(m_1\mid z)$
at different redshifts. Solid curves show the median across prior draws and shaded
regions indicate the 90\% credible interval.}
\label{fig:prior_predictive_m1}
\end{figure}

Finally, we compute the implied prior predictive distribution of the primary-mass
density $p(m_1\mid z)$ by drawing hyperparameters from the prior; Figure~\ref{fig:prior_predictive_m1} shows the resulting prior predictive bands
at representative redshifts.

\section{Hierarchical inference framework}
\label{app:stats}

In this Appendix we provide the full mathematical formulation of the
hierarchical Bayesian inference framework summarized in
Sec.~\ref{sec:stats}.

\paragraph{Hierarchical posterior}

We infer the hyperparameters $\Lambda$ describing the BBH population
and cosmology using a hierarchical Bayesian framework
\cite{Mandel:2018mve}. The data for event $i$ are the strain data, denoted $d_i$, compressed into posterior samples $p(\theta_{D,i}|d_i)$ derived from them under a reference prior $\pi_{\rm PE}(\theta_D)$ and the assumption of statistical independence.

Given a population model $p_{\rm pop}(\theta|\Lambda)$ over source-frame
parameters $\theta$ (masses, spins, redshift), the posterior for the
full catalog of $N_{\rm obs}$ events is
\begin{equation}
\begin{split}
p(\{\theta_{i}\}, \Lambda \mid \{d_i\})
\propto \pi(\Lambda) \,
\xi(\Lambda)^{-N_{\rm obs}}
\prod_{i=1}^{N_{\rm obs}}
\frac{p(\theta_{D,i}|d_i)}{\pi_{\rm PE}(\theta_{D,i})}
\\
\times
p_{\rm pop}\!\left(\theta_i(\theta_{D,i};\Lambda_c)\mid\Lambda\right)
J^{-1}\!\left(\theta(\theta_{D,i};\Lambda_c),\Lambda_c\right),
\end{split}
\label{eq:hier_like_basic_full}
\end{equation}
where the overall merger-rate normalization has been analytically
marginalized over with a scale--invariant prior.

Here $\theta_D$ denotes the detector-frame parameters used in the
single-event parameter estimation, while $\theta$ denotes the
corresponding source-frame parameters. The two sets of parameters are
related by a deterministic mapping
$\theta=\theta(\theta_D;\Lambda_c)$ determined by the cosmological
distance--redshift relation and the conversion between detector-
and source-frame masses.

The factor
\[
J(\theta;\Lambda_c)=\left|\frac{\partial\theta_D}{\partial\theta}\right|
\]
is the Jacobian associated with this change of variables.

Parameters entering the waveform but not explicitly included in
$p_{\rm pop}$ are effectively assumed to follow the reference prior
used in the single-event analysis.

\paragraph{Selection effects}

The term $\xi(\Lambda)$ in Eq.~(\ref{eq:hier_like_basic_app})
represents the detection efficiency, defined as the fraction of
sources drawn from the population model that would be detected by
the search pipeline.
We estimate $\xi(\Lambda)$ using Monte Carlo integration over
simulated injections \cite{Tiwari:2017ndi,Farr:2019rap}. If the
injections $\theta_{D,k}$ are drawn from an injection distribution
$\pi_{\rm inj}(\theta_D)$, the estimator reads
\begin{equation}
\hat{\xi}(\Lambda)=
\frac{1}{N_{\rm inj}}
\sum_{k=1}^{N_{\rm det,inj}}
\frac{
p_{\rm pop}(\theta_k(\theta_{D,k};\Lambda_c)|\Lambda)
}{
\pi_{\rm inj}(\theta_{D,k})\,
J(\theta(\theta_{D,k};\Lambda_c),\Lambda_c)
}.
\label{eq:alpha_inj_app}
\end{equation}

The sum runs over injections recovered by the search pipeline.

\paragraph{Full likelihood evaluation.} The package
\pymcpop  allows sampling the full hierarchical posterior in Eq.~\ref{eq:hier_like_basic_full} in the parameter space $\{ \{\theta_i\}, \Lambda \}$.  A continuous approximation of $p(\theta_{D,i}|d_i)$ is obtained via a Gaussian mixture model, with the number of components selected by minimizing the Bayesian information criterion and used for inference. We refer to Ref.~\cite{Mancarella:2025uat} for a detailed description and to Ref.~\cite{Tagliazucchi:2026dpr} for a blinded mock data challenge where this strategy is shown to be perfectly equivalent to other likelihood evaluation procedures. In this work we use the full sampling scheme when analyzing simulated data.

\paragraph{Posterior-sample reweighting}
In order to simplify the problem, it is customary to analytically marginalize the posterior in Eq.~(\ref{eq:hier_like_basic_full}) over single--event parameters. 
We use this approach to analyze GWTC-4.0 data.
In our production runs we downsample to 5000 PE samples per event to evaluate the per-event Monte Carlo integrals, and we verified that this number is sufficient to guarantee numerical accuracy according to our requirements (see below).
In this case, we evaluate 
\begin{equation}
\begin{split}
p(\Lambda \mid \{d_i\})
\propto \pi(\Lambda) \,
\xi(\Lambda)^{-N_{\rm obs}}
\prod_{i=1}^{N_{\rm obs}} \!\int \dd \theta_{D, i}\, p(d_i|\theta_{D, i})
\\
\times
p_{\rm pop}\!\left(\theta(\theta_{D, i};\Lambda_c)\mid\Lambda\right)
J^{-1}\!\left(\theta(\theta_{D, i};\Lambda_c),\Lambda_c\right)
\end{split}
\label{eq:hier_like_basic_app}
\end{equation}

The per-event integrals in Eq.~(\ref{eq:hier_like_basic_app})
are evaluated by Monte Carlo integration using posterior samples
generated with the reference prior $\pi_{\rm PE}(\theta_D)$ in the
event-level analysis.
Let $\{\theta_{ij}\}_{j=1}^{n_i}$ denote the posterior samples for
event $i$. The integral can then be approximated as
\begin{equation}
\begin{split}
&\int d\theta_{D, i} \,
p(d_i|\theta_{D, i})
p_{\rm pop}(\theta_i(\theta_{D, i};\Lambda_c)|\Lambda)
J^{-1}(\theta_i(\theta_{D, i};\Lambda_c),\Lambda_c)
\\
&\approx
\frac{1}{n_i}
\sum_{j=1}^{n_i}
\frac{
p_{\rm pop}(\theta_{ij}(\theta_{D,ij};\Lambda_c)|\Lambda)
}{
\pi_{\rm PE}(\theta_{D, ij})\,
J(\theta_{ij}(\theta_{D,ij};\Lambda_c),\Lambda_c)
} \equiv \hat{\mathcal{L}}_i(\Lambda).
\end{split}
\label{eq:reweight_event_app_integral}
\end{equation}

\paragraph{Monte Carlo uncertainty}

Monte Carlo sums introduce a stochastic
error in the likelihood estimator
\cite{Farr:2019rap,Essick:2022ojx,Talbot:2023pex,Heinzel:2025ogf}.

Following Refs.~\cite{Talbot:2023pex,Heinzel:2025ogf}, we estimate the
variance of the log-likelihood as
\begin{equation}
\sigma^2_{\ln\hat{\mathcal{L}}}(\Lambda)
\simeq
\sum_{i=1}^{N_{\rm obs}}
\frac{\sigma^2_{\hat{\mathcal{L}}_i}}{\hat{\mathcal{L}}_i^2}
+
N_{\rm obs}^2
\frac{\sigma^2_{\hat{\xi}}}{\hat{\xi}^2},
\end{equation}
where $\hat{\mathcal{L}}_i(\Lambda)$ is the Monte Carlo estimator of the $i$th event integral in Eq.~(\ref{eq:reweight_event_app_integral}), $\sigma^2_{\hat{\mathcal{L}}_i}(\Lambda)$ is its Monte Carlo variance, and $\hat{\xi}(\Lambda)$ and $\sigma^2_{\hat{\xi}}(\Lambda)$ are the injection estimator in Eq.~(\ref{eq:alpha_inj_app}) and its Monte Carlo variance, respectively. When the full likelihood in Eq.~\ref{eq:hier_like_basic_full} is used, only the second term contributes.
Large Monte Carlo fluctuations can bias population inference;
therefore we impose the standard requirement~\cite{Talbot:2023pex,Heinzel:2025ogf,LIGOScientific:2025pvj}
\[
\sigma^2_{\ln\hat{\mathcal{L}}}(\Lambda)<1
\]
and penalize points violating this criterion.

\section{Posterior predictive checks}
\label{app:ppc}

\begin{figure*}
  \centering
      \includegraphics[width=.8\textwidth]{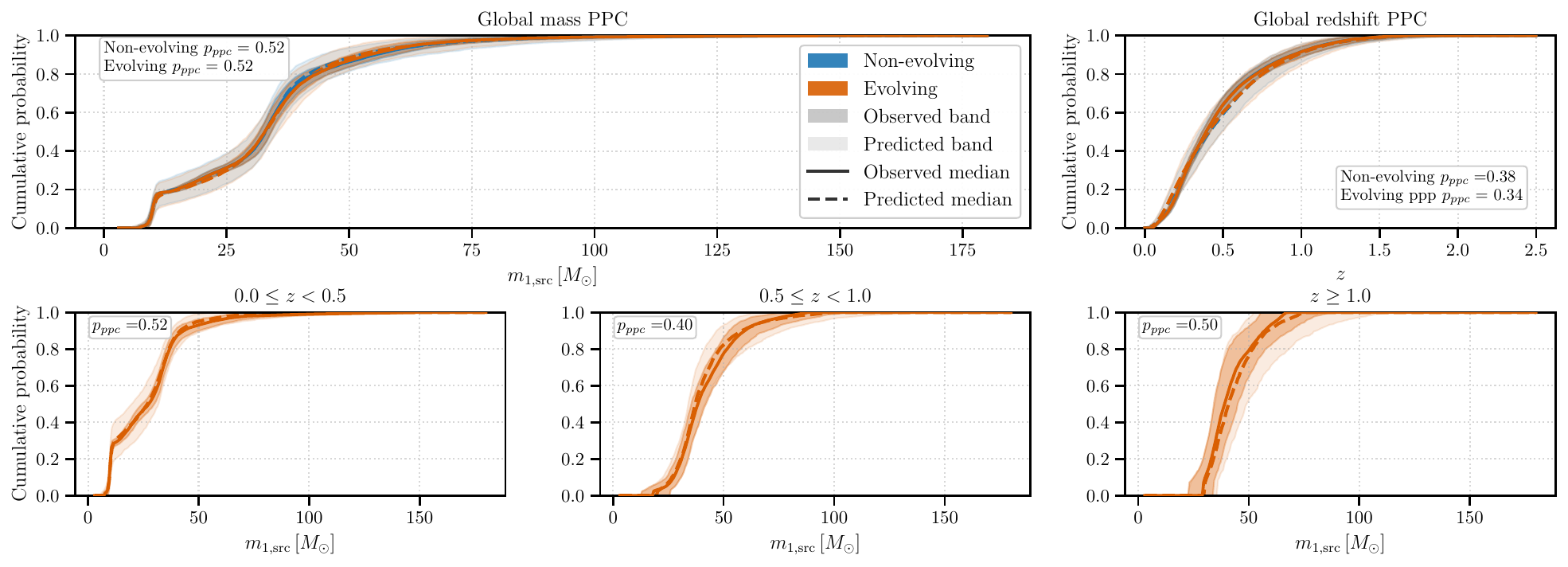}
  \caption{ Posterior predictive checks for the cosmology--fixed analysis. }
  \label{fig:ppc_all_astro}
\end{figure*}

\begin{figure*}
  \centering
      \includegraphics[width=.8\textwidth]{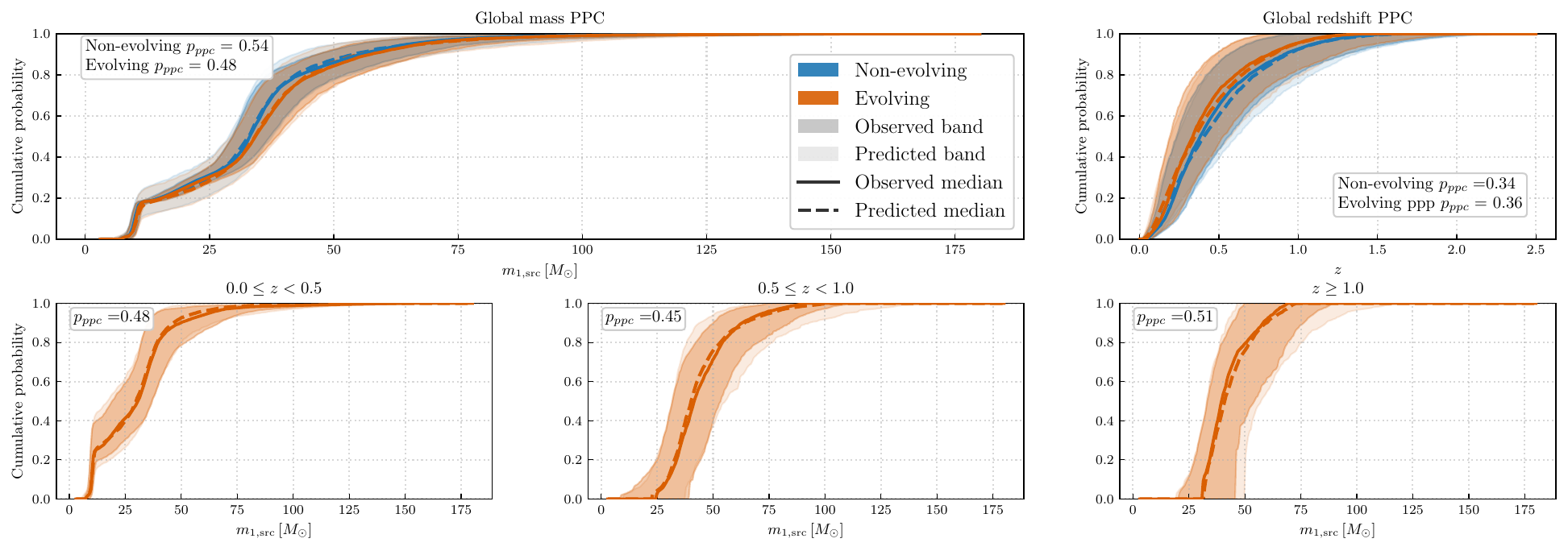}
  \caption{ Posterior predictive checks for the cosmology--free analysis. 
  }
  \label{fig:ppc_all_cosmo}
\end{figure*}

Figure~\ref{fig:ppc_all_astro} shows the posterior predictive distributions for both the non-evolving mass model and the redshift-evolving model in the cosmology--fixed case. The shaded bands represent the $90\%$ predictive intervals of the cumulative distributions, while the lines show the median predicted curves.
Figure~\ref{fig:ppc_all_cosmo} shows the same for the cosmology--free case.
In the cosmology--free case, we obtain posterior predictive $p$-values nearly identical to the cosmology--fixed case, namely 
$\bar p_{\rm ppc}(m_1) \approx 0.96$ ($\bar p_{\rm ppc}(m_1) \approx 0.92$ ),
$\bar p_{\rm ppc}(z) \approx 0.68$ ($\bar p_{\rm ppc}(z) \approx 0.72$),
for the redshift--evolving (non--evolving) model. 
The posterior predictive symmetric $p$-values for the redshift--evolving model with free cosmology are $0.96$, $0.9$, and $0.98$ for the bins $z<0.5$, $0.5\le z<1$, and $z\ge1$.


\section{Mass spectrum constraints with varying cosmology}
\label{app:cosmo_evol}

\begin{figure*}[t]
\centering
\includegraphics[width=0.49\textwidth]{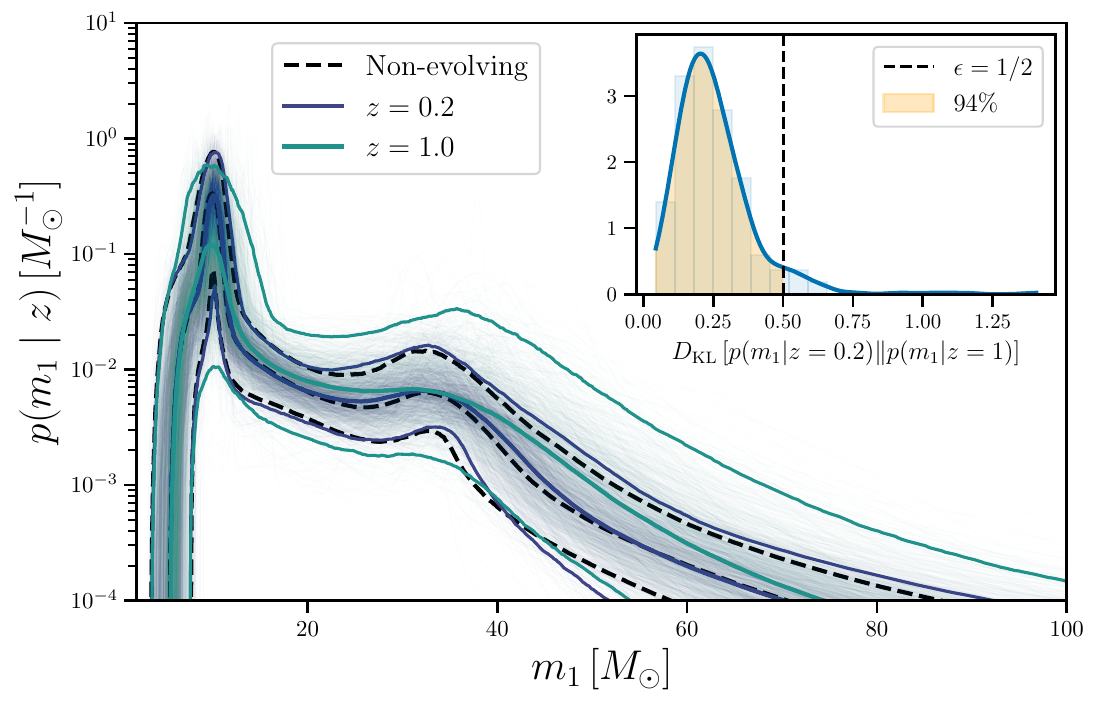}
\hfill
\includegraphics[width=0.49\textwidth]{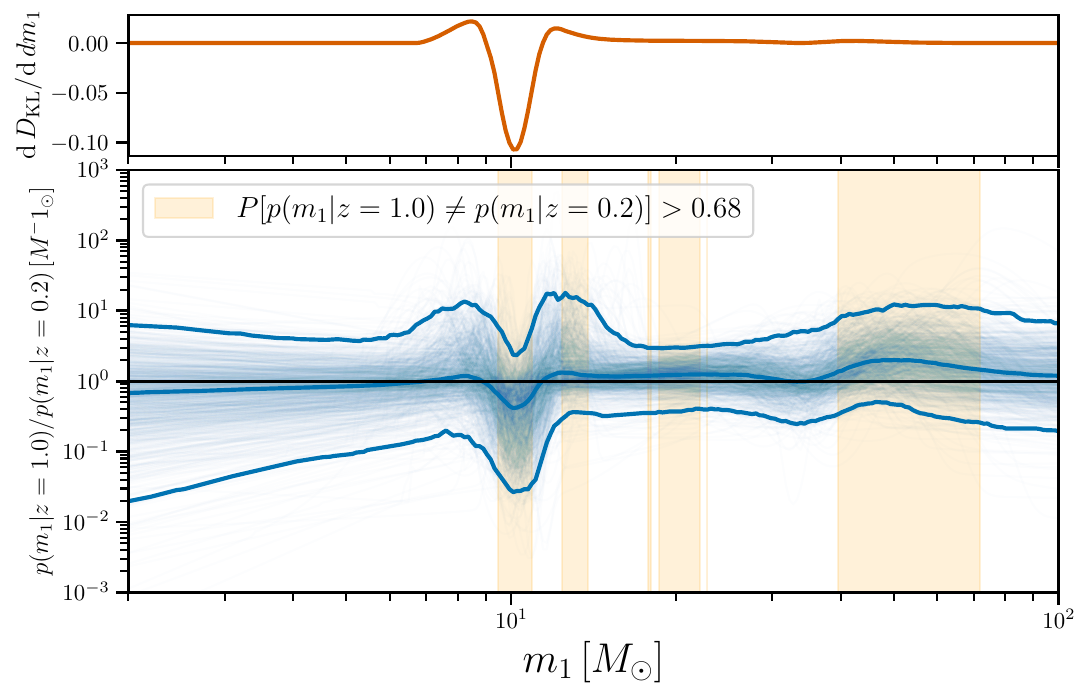}
\caption{Same as Fig.\ref{fig:pm1_marginal_at2_redshifts} but with varying cosmology.
}
\label{fig:pm1_marginal_at2_redshifts_cosmo}
\end{figure*}

\begin{figure*}[t]
\centering
\includegraphics[width=0.24\linewidth]{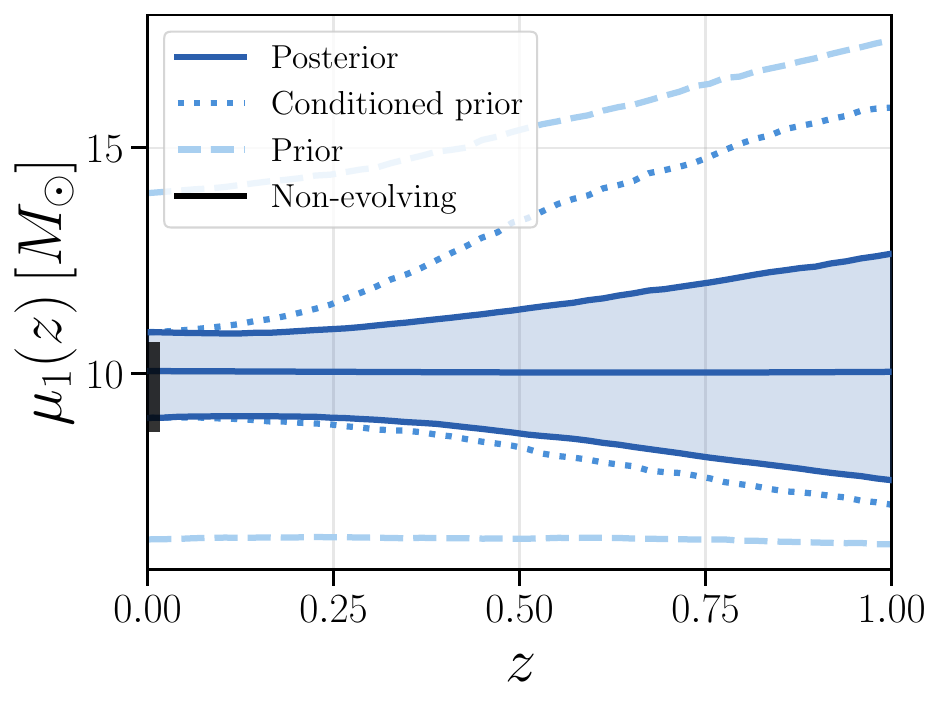}
\hfill
\includegraphics[width=0.24\linewidth]{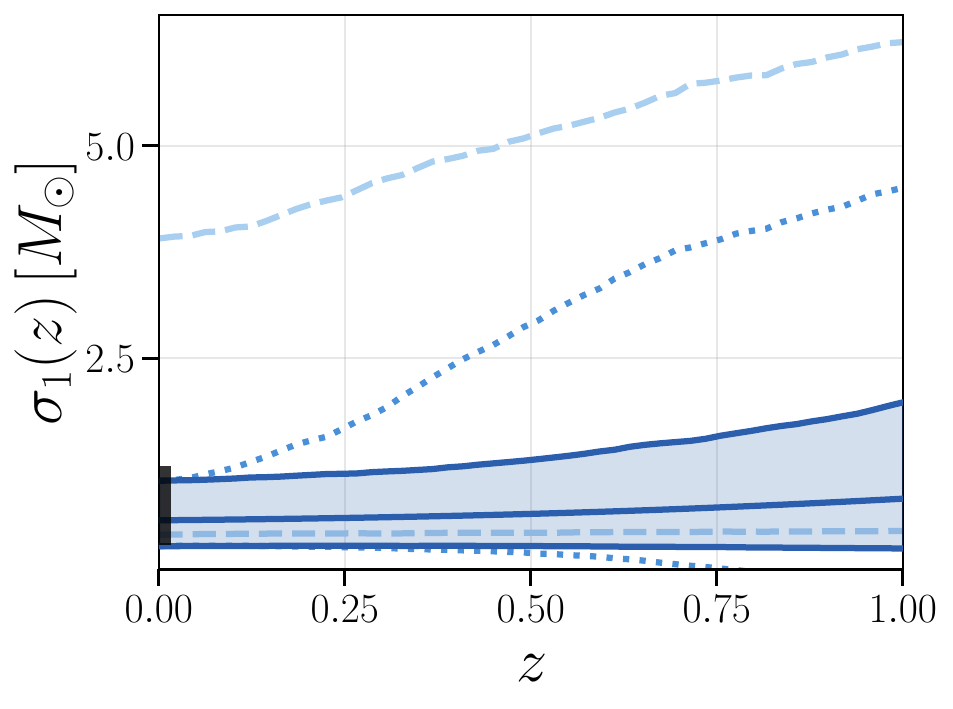}
\hfill
\includegraphics[width=0.24\linewidth]{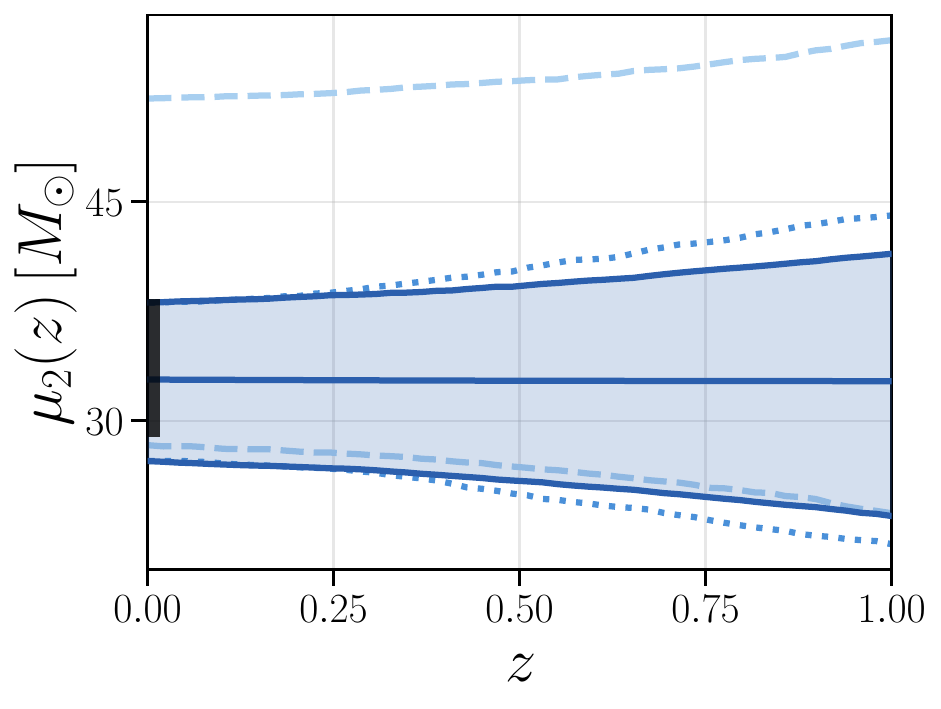}
\hfill
\includegraphics[width=0.24\linewidth]{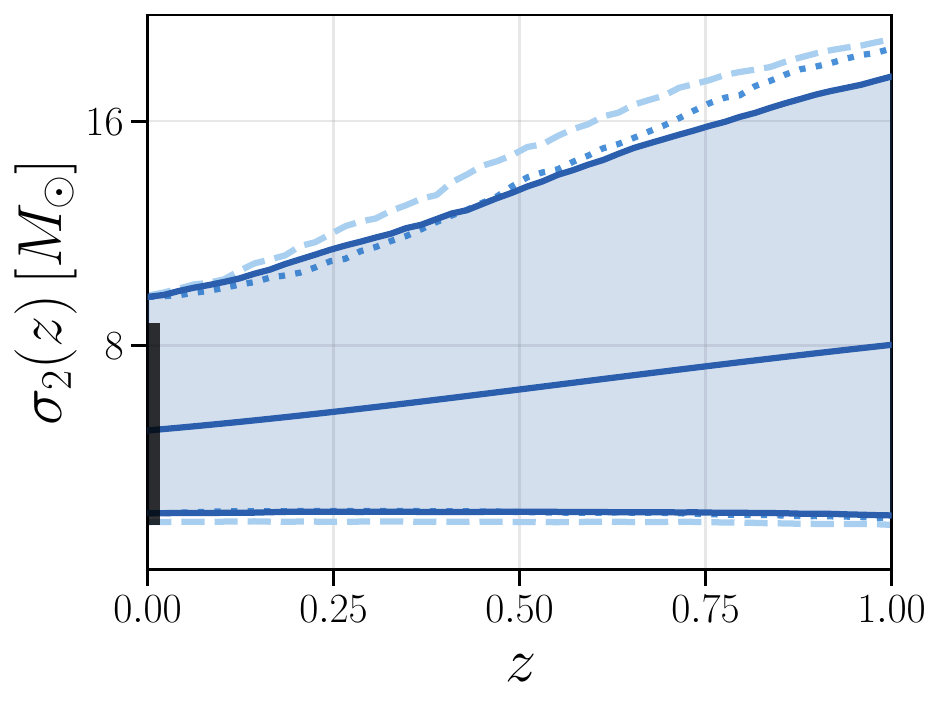}
\caption{Same as Fig.\ref{fig:param_evo} but with varying cosmology.}
\label{fig:param_evo_cosmo}
\end{figure*}

In this Appendix we report figures related to constraints on the mass spectrum (Fig.~\ref{fig:pm1_marginal_at2_redshifts_cosmo}) and on specific hyperparameters (Fig.~\ref{fig:param_evo_cosmo}). 
Results are analog to the cosmology--fixed scenario, although with slightly broader uncertainties.

\end{document}